\documentclass{raa}  
\usepackage{graphicx,times}
\usepackage{natbib}
\usepackage{amssymb,amsmath}
\bibpunct{(}{)}{;}{a}{}{,}
\usepackage{soul,xcolor}
\setstcolor{red}

\usepackage[a4paper=true,pagebackref=true]{hyperref}
\hypersetup{pdftitle = Polarization Observations of NGC~315 with FAST, pdfauthor = My name, pdfsubject= The subject, pdfkeywords = NGC~315 magnetic fields polarization}
\hypersetup{colorlinks = true, linkcolor = green, anchorcolor = red, citecolor = blue, filecolor = red, pagecolor = red, urlcolor = red}

\begin{document}

   \title{Polarization Observations of NGC~315 with FAST}

 \volnopage{Vol.0 (20xx) No.0, 000--000}
   \setcounter{page}{1}

   \author{Wenjun Zhang\inst{1}, Xiaohui Sun\inst{1}, 
   Jie Wang\inst{2}, 
   Xuzhi Hu\inst{1}, 
   Shengtao Wang\inst{1} 
   }

   \institute{School of Physics and Astronomy, Yunnan University, Kunming 650500, China; {\it wjz.astronomy@foxmail.com, xhsun@ynu.edu.cn}\\
        \and
             National Astronomical Observatories, Chinese Academy of Sciences, Beijing 100101, China\\
\vs \no
   {\small Received 20XX Month Day; accepted 20XX Month Day}
}

\abstract{We present new continuum and polarization observations of radio galaxy NGC~315 obtained with the Five-hundred-meter Aperture Spherical radio Telescope (FAST) over 1.0--1.5~GHz. The FAST data detect diffuse large-scale polarized emission with a noise level of 0.065~mJy~beam$^{-1}$ at resolution of $4\arcmin$. We derived a rotation measure (RM) map with an error of 1.1~rad~m$^{-2}$ including systematic error. The intrinsic RM map was obtained after subtracting the Galactic foreground contribution. We estimated a total magnetic field strength of $\lesssim1$~$\mu$G across the source and a thermal electron density of $\sim10^{-4}$~cm$^{-3}$ toward the northwest lobe. The fractional polarization analysis suggests a random magnetic-field component over a large area toward the northwest lobe. 
}

   \authorrunning{Zhang et al.}            %author_head in even pages
   \titlerunning{NGC~315 with FAST}  % title_head in odd pages
   \maketitle

%________________________________________________ sections below
% 

\section{Introduction}           %% first-level sections will be auto-capitalized
\label{sect:intro}
Fanaroff-Riley type I (FR~I) radio galaxies are characterized by their bright nuclear radio emission from the active galactic nucleus (AGN). The radio emission decreases with distance from the core that is usually accompanied by two-sided jets and extended lobe structures \citep{Bicknell+1995}. At the jet termini, interactions with the surrounding intergalactic medium (IGM) frequently give rise to lobes with diverse morphologies. These galaxies are generally found in groups or clusters \citep{Tasse2018}.

FR I radio galaxies often exhibit strong radio emission, making it relatively easy to observe their linear polarization and Faraday rotation measure (RM) and to study their magnetic field. The magnetic field can probe the interaction between galaxies and the IGM. The RM associated with FR~I radio galaxies is predominantly contributed by foreground material. After subtracting the foreground contribution, significant fluctuations in the internal RM are still evident across different regions of FR~I radio galaxies \citep[e.g.][]{Wezgowiec2024}. Starting from the core, the radio jet gradually loses energy during propagation, transitioning from a well-collimated structure to a more extended and diffuse morphology \citep{Fomalont+1980}. As the jet propagates through the IGM, ram pressure compresses the magnetic field, thereby altering its distribution \citep{Klein1994}. Studies of magnetic fields and RM variations in radio galaxies are therefore essential for understanding the interaction between galaxies and the IGM.

NGC~315 was first mapped by \citet{Bridle1976} with the Arecibo telescope at 0.43, 1.41 and 2.38~GHz, revealing a giant radio source with a distinctive Z-shaped morphology. Their maps showed a pronounced large-scale asymmetry between the northwest and southeast sides of the source, with the northwest emission appearing more extended and strongly bent. Follow-up Westerbork Synthesis Radio Telescope (WSRT) observations at 49 and 21~cm presented the total intensities of the jets and extended lobes in greater detail and established NGC~315 as one of the best-studied very large radio galaxies \citep{Willis1981}. Subsequent continuum studies, including WSRT observations at 92 and 49~cm and Effelsberg 100-m telescope observations at 2.8~cm, further demonstrated the presence of diffuse large-scale emission and clarified the complex bending structure on the northwest side \citep{Mack1997,Klein1994}.

Polarization studies of NGC~315 have examined both its large-scale radio structure and the inner northwest jet. Using WSRT observations at 49 and 21~cm, \citet{Willis1981} presented the polarized intensities of the jets and extended lobes and showed that the fractional polarization generally increases with distance from the core along the jets, implying a progressively more ordered magnetic field. They further showed that the dominant magnetic-field component in both the main and counter jets is perpendicular to the jet axis over most of their lengths. Higher-resolution Very Large Array (VLA) observations at 6~cm provided a more detailed view of the inner northwest jet, where \citet{Fomalont+1980} found that the fractional polarization rises from nearly zero near the core to about 14\% further along the jet. Combining the earlier WSRT polarization data with 11~cm measurements, \citet{Fomalont+1980} confirmed, under an adopted average RM, that the projected magnetic field changes from predominantly parallel to perpendicular to the jet axis.

The Faraday rotation in NGC~315 has also been studied. \citet{Willis1981} derived RM toward the jets and lobes using WSRT observations at 49 and 21~cm and found RM variations of up to about 13~rad~m$^{-2}$ across the source. Note that RMs were constrained to $-80\pm8$~rad~m$^{-2}$ to avoid the $n\pi$ ambiguity with only two frequencies. Using VLA observations at five frequencies between 1.365 and 5~GHz, \citet{Laing2006} derived detailed RM distributions only along the jets, finding values typically between $-90\,\mathrm{rad\,m^{-2}}$ and $-60\,\mathrm{rad\,m^{-2}}$, forming a linear gradient along the jet axis. They concluded that both the mean RM and the gradient are dominated by the Galactic foreground, whereas the residual fluctuations, with amplitudes of about $2\,\mathrm{rad\,m^{-2}}$ on scales of 10--100 arcsec, are likely associated with a surrounding hot halo. More recent Very Long Baseline Interferometry (VLBI) studies have focused on parsec-scale jet properties in the nuclear region \citep{Ricci2022}, where no polarized emission has been detected over a wide frequency range from 1 to 43~GHz \citep{Park+2021b}.

%More recent Very Long Baseline Interferometry (VLBI) studies have concentrated on parsec-scale jet properties, using polarization and depolarization behavior to infer the magnetic-field structure in the nuclear region \citep{Park2021,Ricci2022}.

The large-scale polarized structure and RM of the outer lobes, particularly the part extending from the bent northwest lobe, remain poorly explored due to missing large-scale flux in previous interferometric observations and insufficient surface-brightness
sensitivity of previous single-dish data. In this paper, we present new continuum and polarization observations of NGC~315 with the Five-hundred-meter Aperture Spherical radio Telescope (FAST) revealing the large-scale structures. The paper is organized as follows. Observation and data processing are presented in Sect.~\ref{sect:obs}, results are presented in Sect.~\ref{sect:res} followed by discussions in Sect.~\ref{sect:dis}, and conclusions are presented in Sect.~\ref{sect:con}.

%have remained poorly explored because of the limited sensitivity to diffuse emission in previous interferometric and single-dish observations.

The redshift of NGC~315 is $z=0.0164$~\citep{Trager+2000}, and the angular scale at the distance of NGC~315 is approximately $0.347\,\mathrm{kpc\,arcsec^{-1}}$ with cosmological parameter $H_0=67.4\,\mathrm{km\,s^{-1}\,Mpc^{-1}}$~\citep{Planck+2020}.

\section{Observations and data reduction}
\label{sect:obs}
NGC~315 was covered in an observation campaign targeting M~31 and its halo with FAST (PI: Jie Wang). The observations were conducted with the DriftWithAngle mode \citep{Jiang2020}. In this mode, the 19-beam receiver is rotated by $23\fdg4$ and fixed at declination, and the sky drifts through the telescope. A drift scan delivers a declination strip of about $22\arcmin$ width in declination with Nyquist sampling. The scan speed is $15\arcsec$~s$^{-1}$. In order to calibrate the system variations, a linearly polarized noise of about 12.5~K is injected into the system for 2~s every 1000~s. The frequency range of the receiving system is 1.0--1.5~GHz, which is divided into 65536 channels. The width of each channel is about 7.63~kHz. The backend outputs four products: auto-correlation and cross-correlation data from orthogonal $X$ and $Y$ feeds, with two for total intensity $I_1$  and $I_2$ , and the other two for Stokes $U$ and $V$. 

Data processing for H\,{\sc I} for the observation campaign has been developed by \citet{Jing+24}, and H\,{\sc I} studies such as the satellite galaxies of M~31 and the Galactic structures have been presented by \citet{Liu+24} and \citet{Liang+25}. For continuum and polarization, we followed the procedure developed by \citet{Sun2021}, \citet{Sun+22} and \citet{Gao+22} to produce images of $I$, $Q$ and $U$, and a study of M~31 was presented by \citet{Zhang+23}.

The data were processed as follows. We first took the median of every 128 adjacent channels. This reduced the number of channels from 65536 to 512, with each channel having a width of 976.56 kHz. Radio frequency interference (RFI) was then identified using a median filter and removed, leaving 287 usable channels covering the frequency range of 1.033--1.464~GHz~(Table~\ref{tab:rmsynth}), with a gap over 1.14--1.29~GHz. Calibration was carried out in two steps: (1) based on injected noise, the temperature scale was established, instrumental polarization leakage and polarization angle were corrected. (2) The flux calibrator 3C 48 was used to convert the temperature in K to the flux density in Jy~beam$^{-1}$. The calibrated drift scans were then combined to produce images of $I$, $Q$, and $U$. The process was repeated for all frequency channels. All images were smoothed to a common resolution of $4\arcmin$ to obtain frequency cubes. 

RM synthesis was applied to the frequency ($\nu$) cubes $Q(\nu)$ and $U(\nu)$ to obtain the Faraday depth ($\phi$) cubes $Q(\phi)$ and $U(\phi)$, because $\mathcal{P}(\lambda^2)\equiv Q(\lambda^2)+iU(\lambda^2)$ and $F(\phi)\equiv Q(\phi)+iU(\phi)$ are a Fourier transform pair~\citep{Brentjens+05}. Here $\lambda$ is the wavelength. The limited frequency bandwidth means a sampling function in the $\lambda^2$ domain, resulting in a point spread function in the $\phi$ domain, which is named RM spread function (RMSF). The full width at half maximum (FWHM) of RMSF, the maximum $\phi$ scale, and the maximum $\phi$ calculated using the formulae by \citet{Brentjens+05} are listed in Table~\ref{tab:rmsynth}. A deconvolution process, called RM clean~\citep{Heald+09}, was applied to derive the true $F(\phi)$. For each pixel, the maximum value of $|F(\phi)|$ was searched and the peak value and the peak location corresponded to the polarized intensity $P$ and RM.  All the above processes were conducted using RM-tools \citep {Purcell2020, VanEck+26}. The Faraday depth range was set to $\pm2000~\mathrm{rad\,m^{-2}}$ with an interval of $13\,\mathrm{rad\,m^{-2}}$ corresponding to $\sim 1/6$ of FWHM of the RMSF. The frequency gap causes large sidelobes in RMSF. RM clean can substantially suppresses the RMSF sidelobes and produce correct RMs, as demonstrated by \cite{Sun2021} using a handful polarized sources.

The residual polarization leakage is less than 1\% after calibration with injected reference noise and shows no dependence on frequency \citep{Sun2021}. After RM synthesis, the residual leakage will appear at $\phi\approx0\,\mathrm{rad\,m^{-2}}$. The systematic RM error due to polarization angle calibration is less than $1\,\mathrm{rad\,m^{-2}}$ \citep{Sun2021}.

\begin{table}
\centering
\caption{Parameters of FAST observation and RM synthesis}
\label{tab:rmsynth}
\begin{tabular}{lc}
\hline
\hline
Parameter & Value \\
\hline
Frequency range & 1.0--1.5~GHz \\
Usable frequency range & 1.033--1.464~GHz \\
Angular resolution & $4\arcmin$ \\
Number of usable channels & 287 \\
Channel width & 976.56~kHz \\
FWHM of RMSF & 81.8~rad~m$^{-2}$ \\
Maximum $\phi$ scale & 75.0~rad~m$^{-2}$ \\
Maximum $|\phi|$& $3.1\times10^{4}$~rad~m$^{-2}$ \\
\hline
\end{tabular}
\end{table}

\section{Results}
\label{sect:res}
\subsection{Total Intensity}
\subsubsection{FAST image}
\begin{figure}[!htbp]
   \centering
   \includegraphics[width=0.95\textwidth]{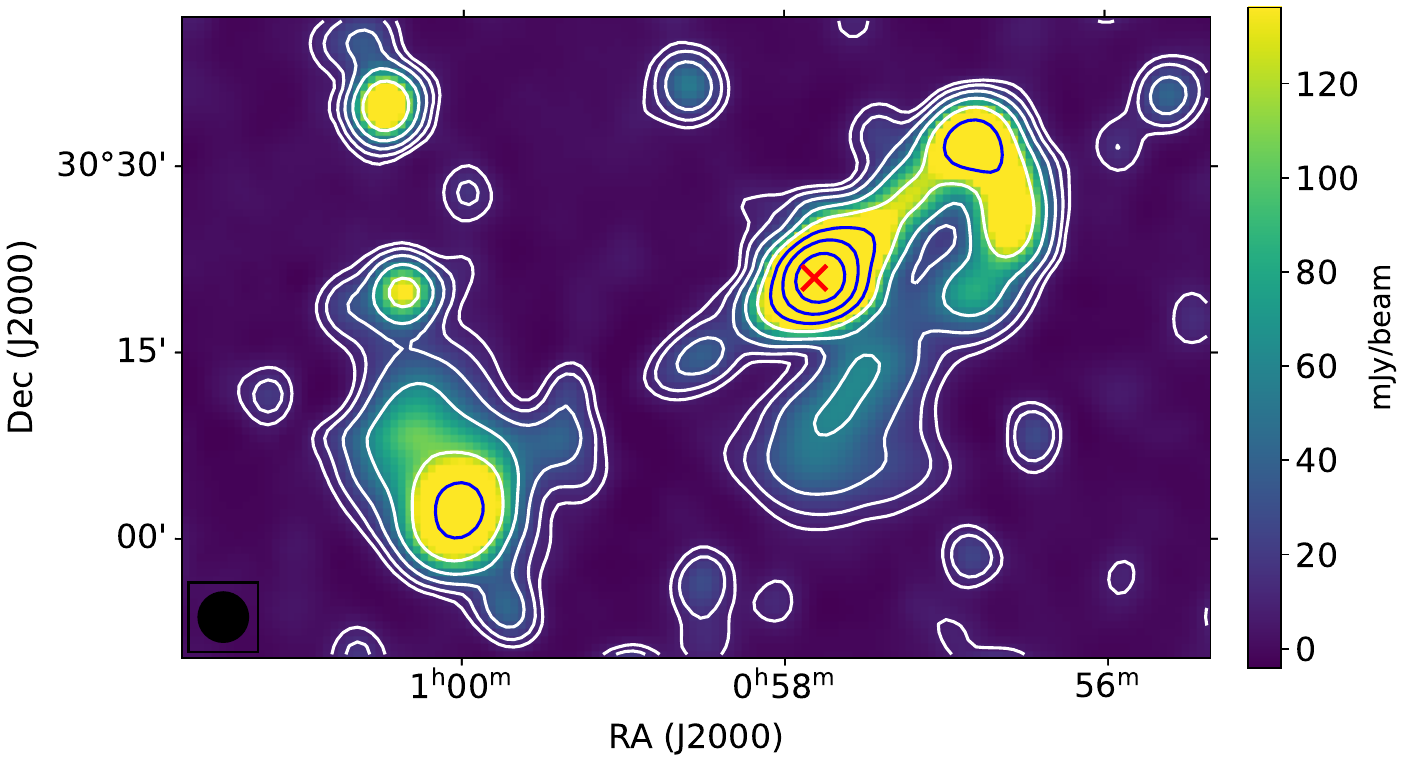}
   \caption{Total intensity image of NGC~315 at 1.248~GHz. The contours represent total intensity levels at $2^n\times 5\sigma_I$ with the rms noise $\sigma_I=1.4$~mJy~beam$^{-1}$ and $n=0,1,2,3....$. The contours overlaid on total intensity peaks are shown in blue. The red cross marks the position of the core. The $4\arcmin$ beam is shown in the lower-left corner.}
   \label{fig:I}
\end{figure}

%wjz_c
We took the median of the $I$ frequency cube and derived the total intensity at 1.248~GHz that is the center of usable frequency range with a resolution of $4\arcmin$, shown in Fig.~\ref{fig:I}. We measured a root mean square (rms) noise level of $\sigma_I=1.4$~mJy~beam$^{-1}$ from areas devoid of strong emission. According to \citet{Condon+1974} and \citet{Meyers+17}, the predicted confusion limit is 2.7 mJy~beam$^{-1}$. However, \citet{Sun+22} measured an rms of 1.3 mJy~beam$^{-1}$, implying that the predicted confusion limit is probably overestimated. Since the integration time of our observations is less than that of \citet{Sun+22}, we have not yet reached the confusion limit. In Fig~\ref{fig:I}, only areas with $I>5\sigma_I$ are shown.

The total intensity image from FAST observations clearly exhibits the large scale structure of NGC~315. The strongest emission is from the core at $({\rm RA, Dec})=(00^{\rm h}57^{\rm m}48^{\rm s}$, $+30\degr21\arcmin08\arcsec)$ determined from VLBI observations by \citet{Charlot+2020}, reaching a level of $\sim$1.5~Jy~beam$^{-1}$. Toward the northwest, the jet extends from the core continuously and terminates at the narrow lobe, and a ``back-flowing'' structure extends from the lobe and is almost parallel to the jet. Toward the southeast, the jet is intermittent, the lobe is wider, and there is no extended structure from the lobe. The general morphology is consistent with previous observations such as those presented by \citet{Mack1997}. 

\subsubsection{Spectral index map}
\label{sect:spectral}
We collected archival total intensity images of NGC~315 at the following frequencies: 144~MHz observed by the LOw-Frequency ARray \citep[LOFAR,][]{Shimwell+2026}, 326 and 609 MHz observed by WSRT and 2700, 4800, and 10700 MHz observed by the Effelsberg 100-m telescope~\citep{Mack1997}. Combining the archival and FAST $I$ images, we obtained the spectral index map by fitting pixels to a power law $S_\nu\propto\nu^\alpha$, where $S_\nu$ is the flux density at frequency $\nu$ and $\alpha$ is the spectral index. All images were convolved to a common angular resolution of $4\farcm35$ and only pixels with flux densities greater than three times the rms noise were used for fitting. The northwest region of NGC~315, which is the main focus of this work has an angular scale of $\sim40\arcmin$, smaller than the recoverable largest angular scales of the LoTSS-DR3 and WSRT observations, approximately $180\arcmin$, $88\arcmin$, and $47\arcmin$ at 144, 326, and 609~MHz. These values were estimated using the shortest baselines of 40~m for LoTSS-DR3 \citep{Shimwell+2026} and 36~m for WSRT \citep{Bridle1976}. However, the full extent of NGC~315 is approximately $1^\circ$, and thus some large-scale flux may be missed at 609~MHz, as noted by \citet{Mack1997}. This may affect the derived spectral indices and fractional polarization at this frequency point. The
  spectral indices, fitted over many frequencies, e.g. Fig.~\ref{fig:appendix_spectra}, are
  unlikely to be strongly affected by a single outlier. We left out the most outside area where weak diffuse emission dominates for fitting, so that the influence of lacking sensitivity for interferometric and previous single-dish observations can be reduced. 

\begin{figure}[!htbp]
    \centering
    
    \begin{subfigure}{\textwidth}
        \centering
        \includegraphics[width=0.85\textwidth]{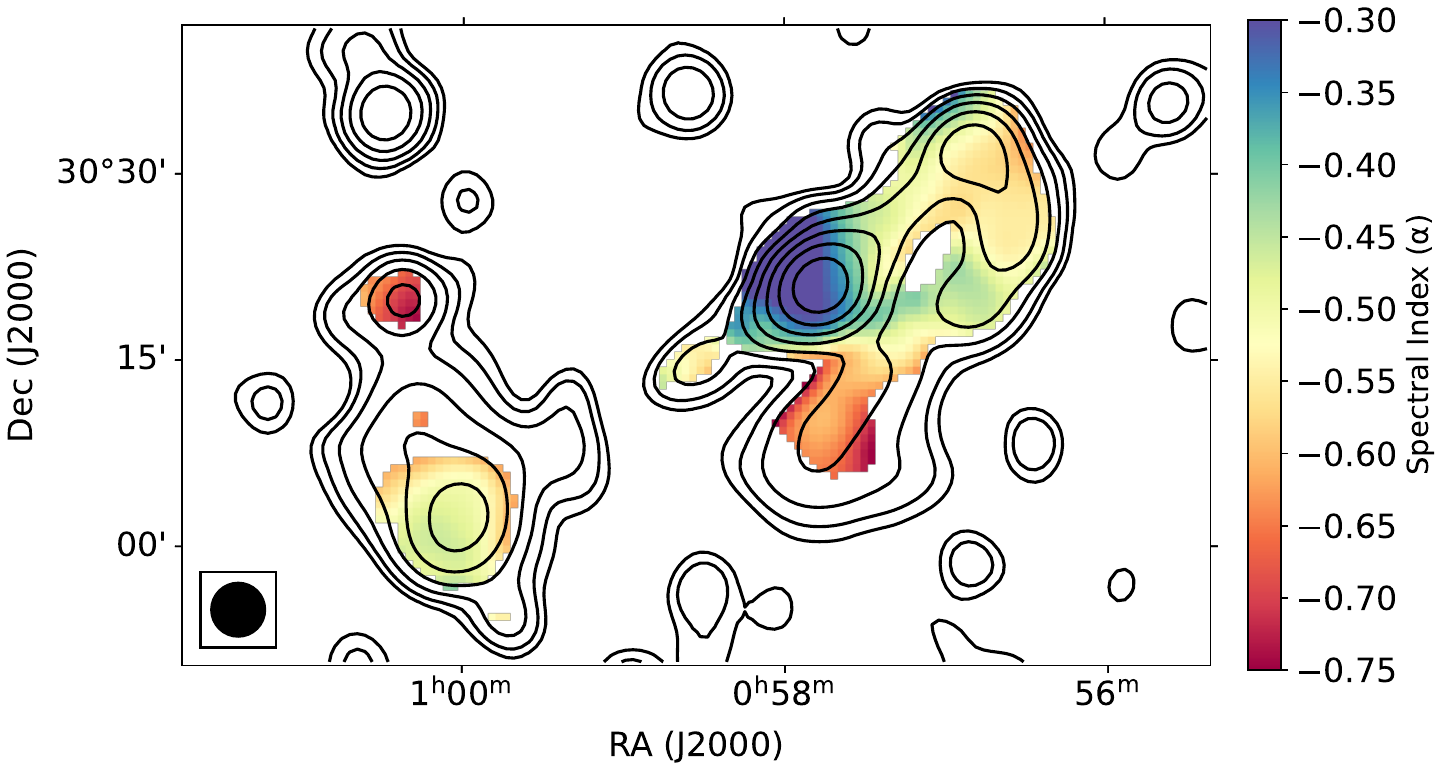}
    \end{subfigure}
    
    \vspace{0.5cm}
    
    \begin{subfigure}{\textwidth}
        \centering
        \includegraphics[width=0.85\textwidth]{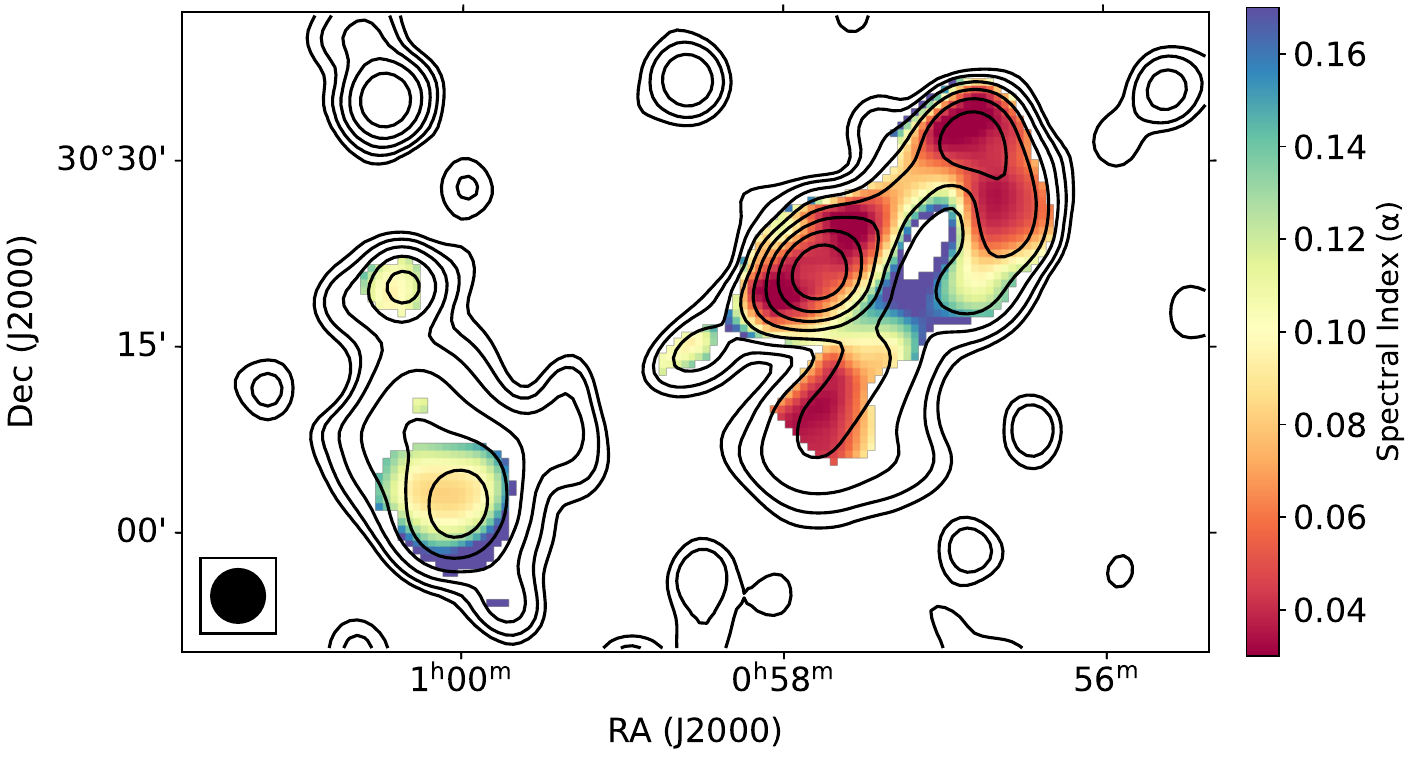}
    \end{subfigure}
    
    \caption{Top panel: Spectral index map of NGC~315 at a resolution of $4\farcm35$. The contours represent the total intensity with levels the same as in Fig.~\ref{fig:I}. Bottom panel: Spectral index uncertainty map with the uncertainties scaled by $\sqrt{\chi^2/\mathrm{dof}}$.}
    \label{fig:spc_map}
\end{figure}

The fitting results for some selected pixels are shown in Fig.~\ref{fig:appendix_spectra}. The derived spectral index map is shown in Fig.~\ref{fig:spc_map} and the uncertainty map is scaled by $\sqrt{\chi^2/\mathrm{dof}}$. The uncertainties are relatively large at some positions, suggesting possible deviations from a single power law over the full frequency range. Such frequency-dependent spectral variations in NGC 315 were also noticed in previous studies \citep{Sohn+2003, Mack1998}. There are no strong point-like sources coincident with the core, jet, and northwest lobe, which can be seen from the LOFAR data \citep{Shimwell+2026} with a high resolution of $20\arcsec$. Therefore, the spectral indices towards these areas are from NGC~315.

The core exhibits a relatively flat spectrum with $\alpha \approx -0.30$. Toward the northwest, the spectral index is $\alpha \approx -0.50$ along the jet, $\alpha \approx -0.55$ across the lobe, and $\alpha \approx -0.63$ in the lobe extension. The differences between these regions are not significant, but they are all steeper than that of the core region. Moving from the core toward the southeast, the spectrum steepens, likewise, along the jet. For the southeast lobe, limited by the sensitivity at higher frequencies, a reliable spectral index cannot be determined over the entire area. The spectral index was obtained only for the central part of the lobe, which is $\alpha \approx -0.50$, similar to that in the northwest lobe. Note that the very steep spectrum toward the north part of the southeast lobe is from the extragalactic source at ${\rm (RA,\, Dec)}={\rm (1^{h}00^{m}23^{s},\, 30\degr20\arcmin06\arcsec)}$.

%The core exhibits a relatively flat spectrum with $\alpha \approx -0.30$. Moving from the core toward the northwest, the spectrum gradually steepens and  the spectral index decreases to $\alpha \approx -0.50$ along the jet; across the lobe, the spectral index remains nearly constant at $\alpha \approx -0.55$; towards the lobe extension region, the spectral index decreases further to $\alpha \approx -0.63$.

\subsection{Polarization}

\subsubsection{FAST Polarized intensity and fractional polarization}
\begin{figure}[!htbp]
   \centering
   \includegraphics[width=0.95\textwidth]{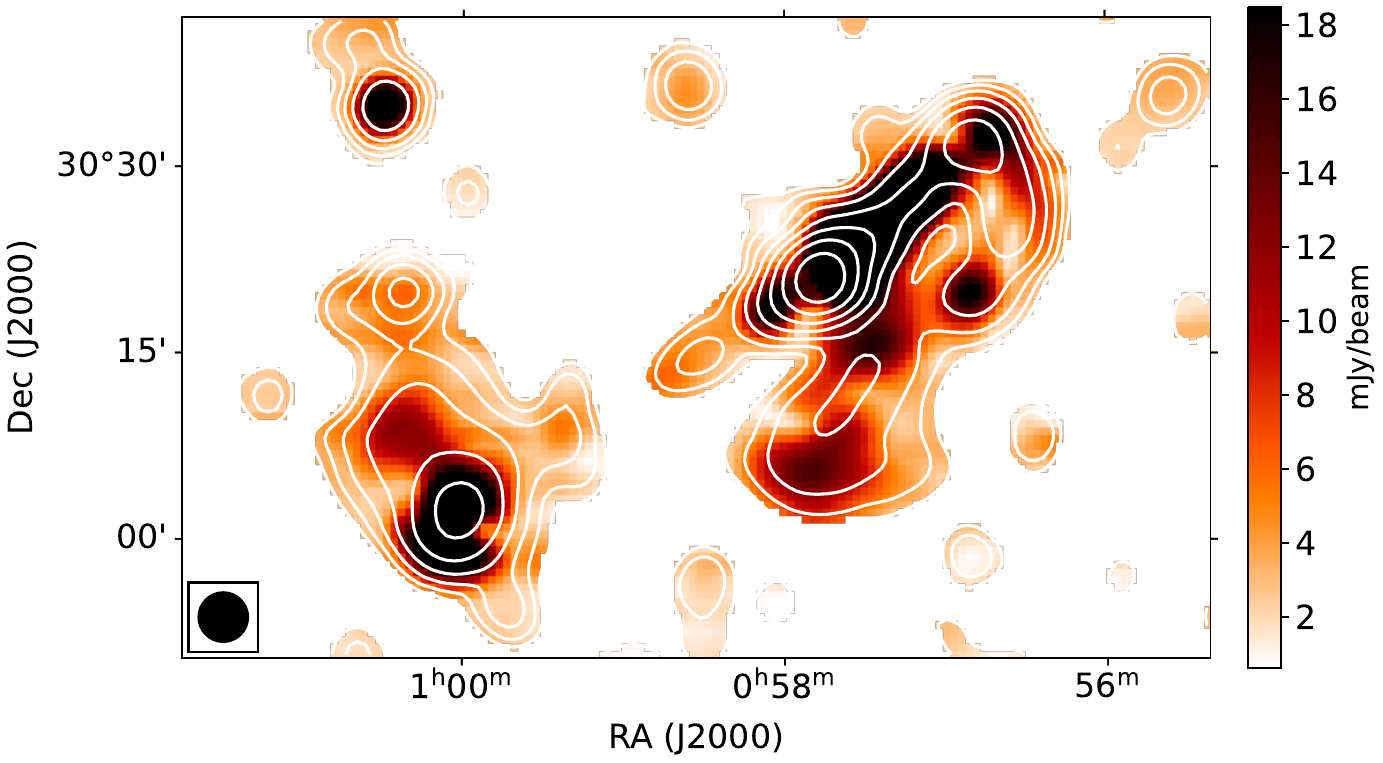}
   \caption{Polarized intensity image of NGC~315 at 1.248~GHz. The contours represent the total intensity levels the same as in Fig.~\ref{fig:I}.}
   \label{fig:PI}
\end{figure}

\begin{figure}[!htbp]
   \centering
   \includegraphics[width=0.95\textwidth]{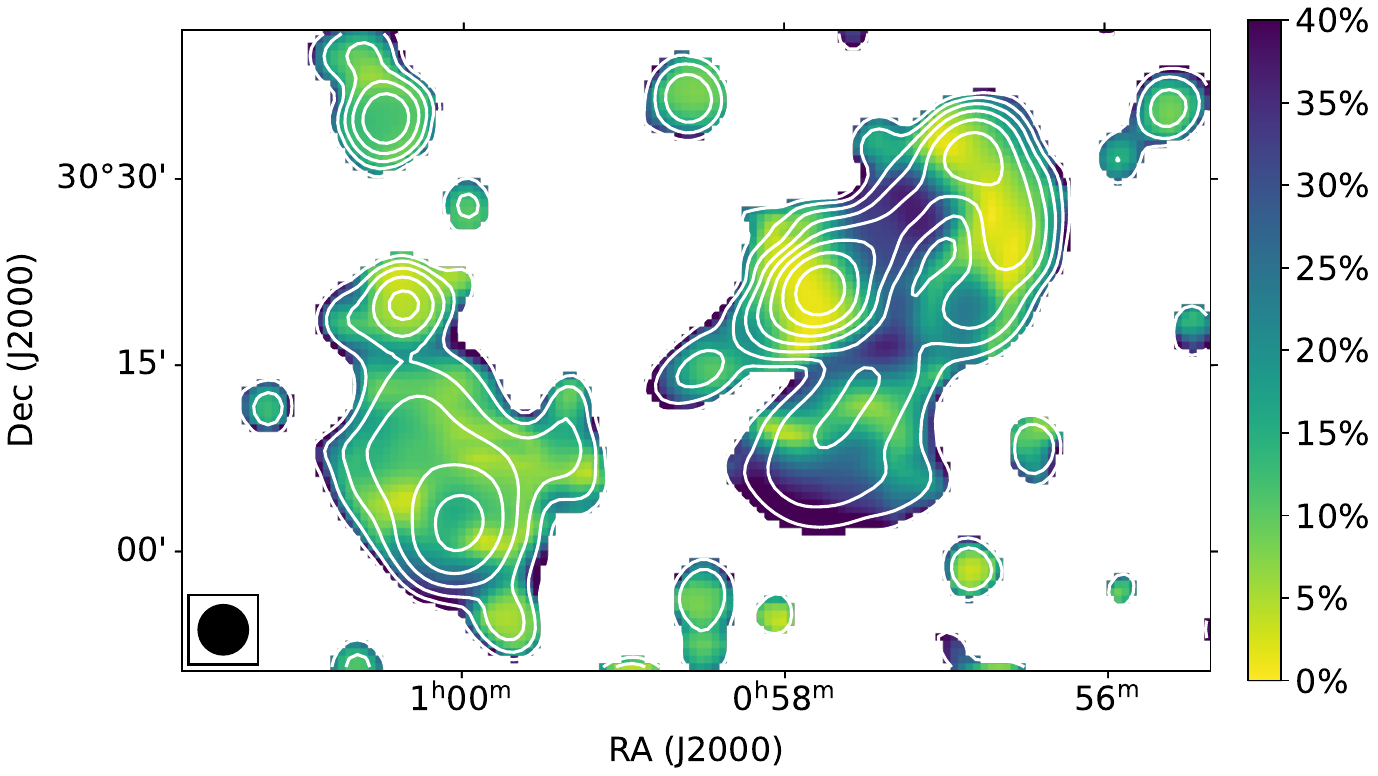}
   \caption{Fractional polarization image of NGC~315 at 1.248~GHz. The contours are the same as in Fig.~\ref{fig:I}.} 
   \label{fig:PD}
\end{figure}

The FAST polarized intensity ($P$) of NGC~315, derived as the peak of $|F(\phi)|$, is shown in Fig.~\ref{fig:PI}. We measured rms noise from $Q(\phi)$ and $U(\phi)$ from Faraday depth cubes in frames with $|\phi|>1000$~rad~m$^{-2}$ where there are no sky signals. The resulting rms noise in $P$ is $\sigma_P=0.065$~mJy~beam$^{-1}$, much smaller than $\sigma_I$ because the confusion noise is much smaller in polarization. In Fig.~\ref{fig:PI}, in order to show the polarized emission from NGC~315, only areas with $I>5\sigma_I$ are displayed. We also derived fractional polarization defined as $p=P/I$ for $P>5\sigma_P$ and $I>5\sigma_I$, shown in Fig.~\ref{fig:PD}.

The core shows very low fractional polarization at the level of $p\approx 3\%$, consistent with high-frequency observations~\citep{Mack1997,Laing2006}. Despite the fractional polarization is low, the RM as shown in Sect.~\ref{sect:RM} deviates largely from 0, indicating that it is unlikely from residual polarization leakage. Toward the northwest, strong polarized intensity persists in the jet with fractional polarization $p\approx 30\%$; fractional polarization decreases to $p\approx5\%$ in the lobe and increases to $p\approx20\%$ in the ``back-flowing'' extension area of the lobe. Toward the southeast, moderate fractional polarization of $p\approx15\%$ was detected in the jet and of $p\approx10\%$ was detected in the lobe.

\subsubsection{RM and magnetic field orientation}
\label{sect:RM}

\begin{figure}[!htbp]
    \centering
    \includegraphics[width=0.95\textwidth]{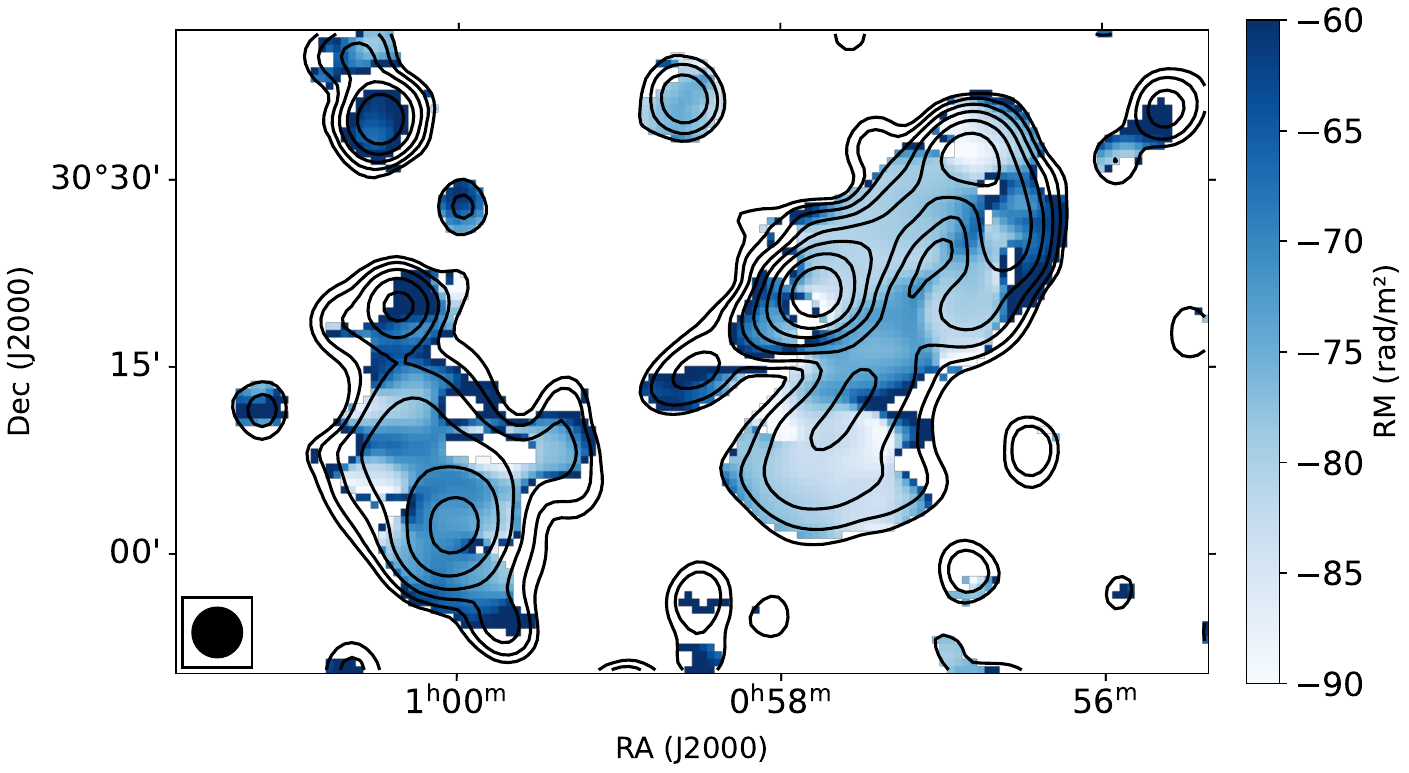}
    \caption{RM image of NGC~315. The contours are the same as in Fig.~\ref{fig:I}.}
    \label{fig:RM_original}
\end{figure}

The RM image of NGC~315, derived as the peak location of $|F(\phi)|$, is shown in Fig.~\ref{fig:RM_original}. The RM error depends on the point spread function in $\phi$ and the signal to noise ratio in $P$~\citep[e.g.][]{VanEck+26}. We estimated a median RM error of 0.5~rad~m$^{-2}$. The total RM observational error including systematics is 1.1~rad~m$^{-2}$.

RM values are all negative, ranging from about $-90$ to $-60\,\mathrm{rad\,m^{-2}}$. There are no large variations of RM across the source. RM is around $-80$~rad~m$^{-2}$ toward the northwest lobe and its extension, and is around $-70$~rad~m$^{-2}$ toward the southeast lobe. 

\begin{figure}[!htbp]
   \centering
   \includegraphics[width=0.95\textwidth]{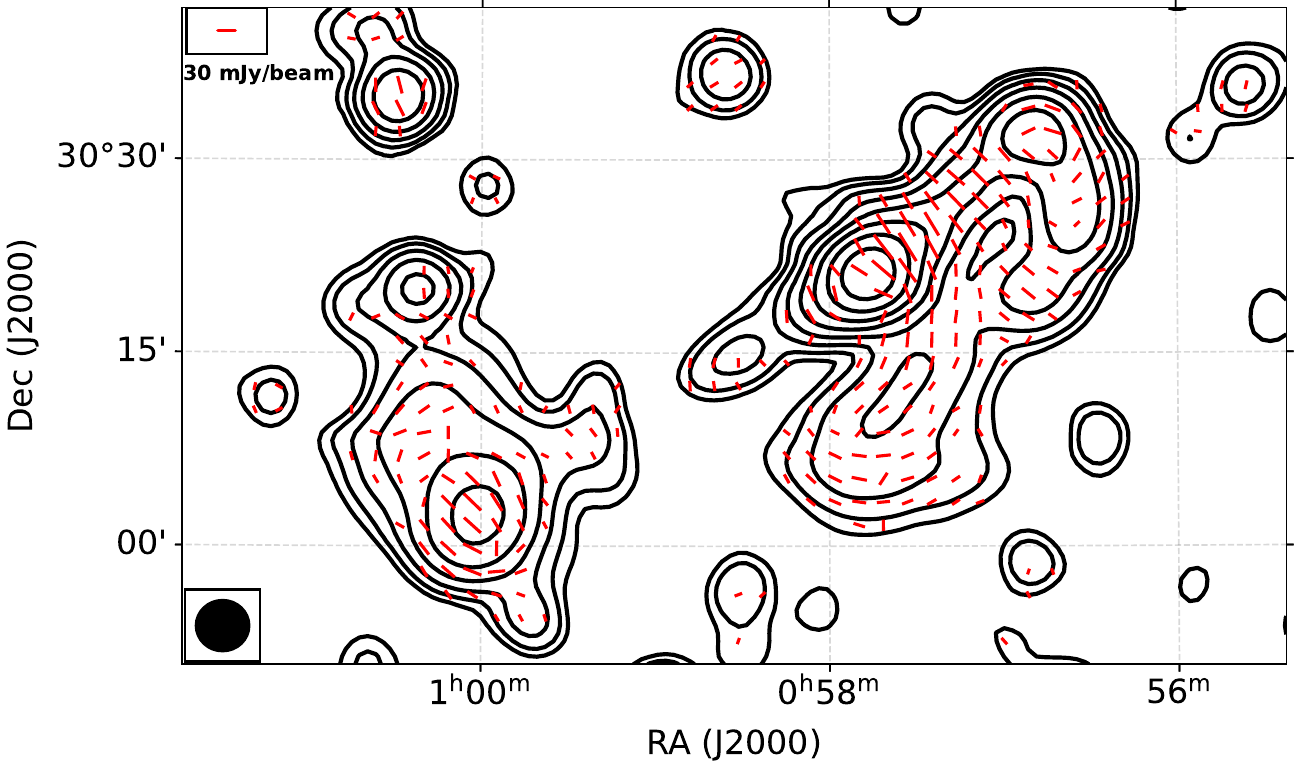}
   \caption{Magnetic-field orientation after RM correction, shown by the bars. The length of the bars is proportional to $\sqrt{P}$. The contours are the same as in Fig.~\ref{fig:I}.} 
   \label{fig:PV}
\end{figure}

We derived the intrinsic polarization angle by correcting for RM and rotated the angle by $90\degr$ to obtain the orientation of the magnetic field, shown in Fig.~\ref{fig:PV}. In the northwest jet, the magnetic field is perpendicular to the jet axis. The magnetic field in the outer parts of the southeast lobe and the northwest lobe extension seems to be loop-like, which was also shown by \citet{Willis1981}. However, this is not definitive.

\section{Discussion}
\label{sect:dis}
\subsection{Magnetic Field and thermal electron density}
\subsubsection{Intrinsic RM}
\begin{figure}[!htbp]
    \centering
    
    \begin{subfigure}{\textwidth}
        \centering
        \includegraphics[width=0.85\textwidth]{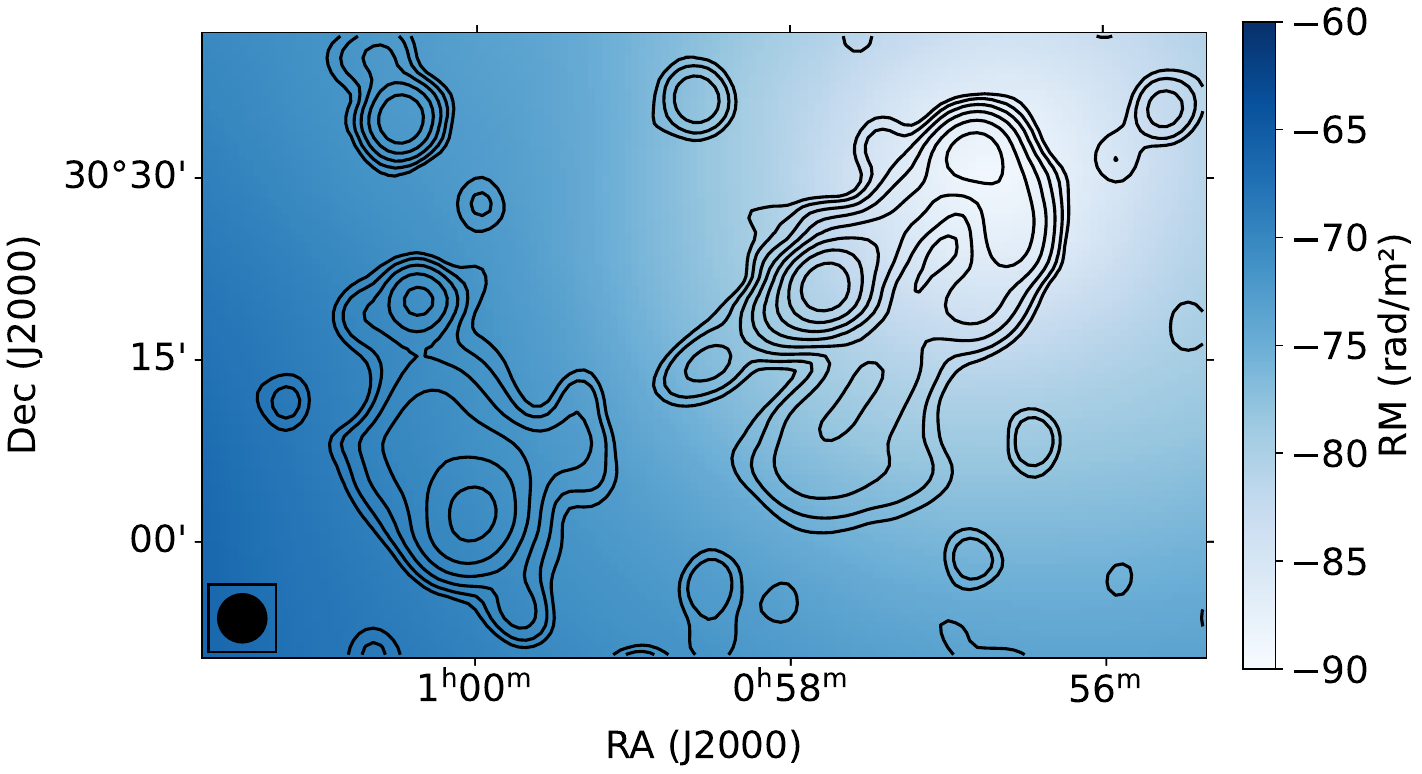}
    \end{subfigure}
    
    \vspace{0.5cm}
    
    \begin{subfigure}{\textwidth}
        \centering
        \includegraphics[width=0.85\textwidth]{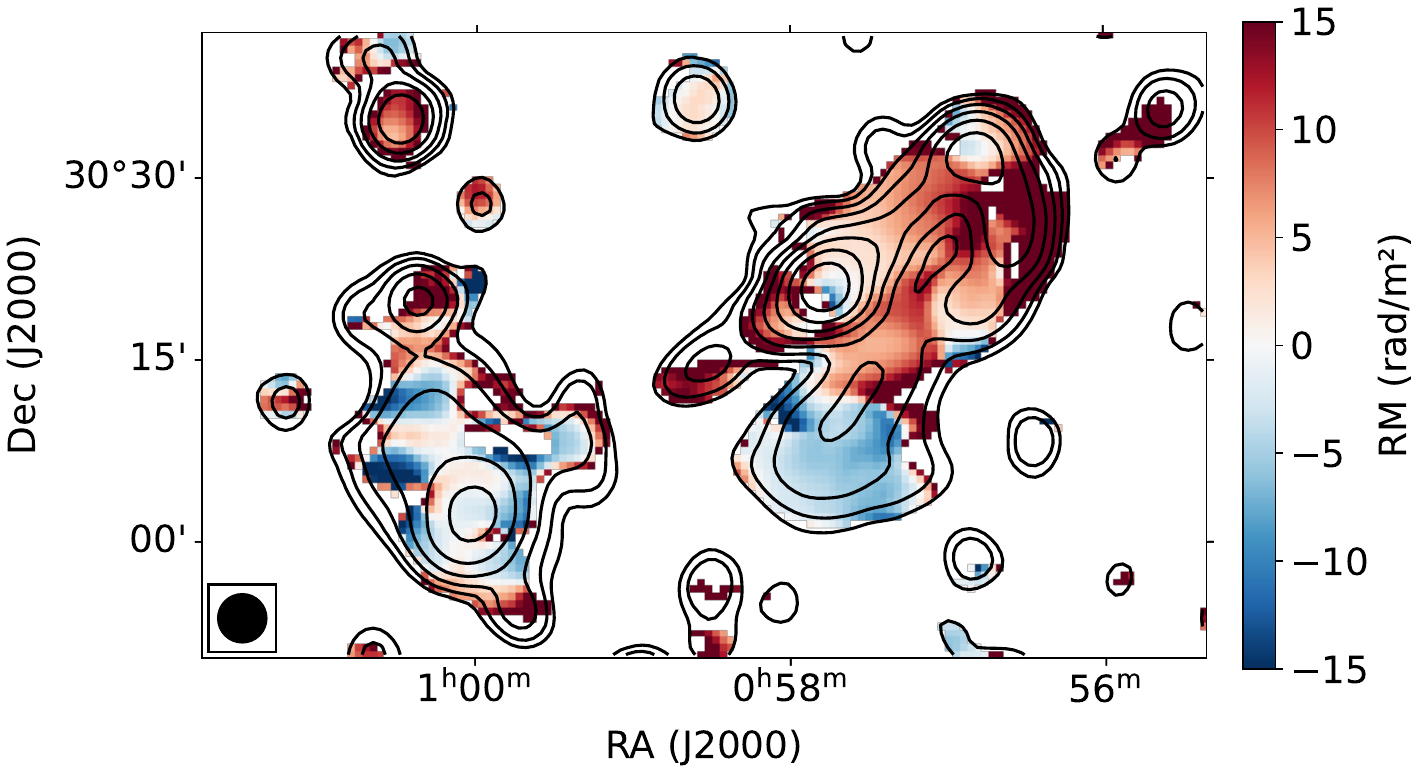}
    \end{subfigure}
    
    \caption{Top panel: Galactic foreground RM from \citet{hutschenreuter2022}; bottom panel: RM of NGC~315 after subtracting Galactic foreground.}
    \label{fig:RM_comparison}
\end{figure}

RMs in Fig.~\ref{fig:RM_original} contain an external contribution from the Galactic foreground and an intrinsic contribution from NGC~315. The Galactic foreground RM was derived by \citet{hutschenreuter2022} using RMs of extragalactic sources, shown in Fig.~\ref{fig:RM_comparison} (top panel). We subtracted the Galactic foreground RM contribution and obtained the intrinsic RM of NGC~315, shown in Fig.~\ref{fig:RM_comparison} (bottom panel). The uncertainty of the foreground RM model is about 9.4~rad~m$^{-2}$, mainly because few extragalactic sources are available to constrain the model in this region~\citep{hutschenreuter2022}. Combining this with the observational error of 1.1~rad~m$^{-2}$ gives a total uncertainty of 9.5~rad~m$^{-2}$ for the intrinsic RM. Note that the intrinsic RM should be multiplied by the factor $(1+z)^2$ to account for the effect of the redshift. For NGC~315, the correction is negligible due to small $z$. Toward the northwest, the intrinsic RM is predominantly positive around $+10$~rad~m$^{-2}$ in the jet and lobe after subtracting the modeled foreground. In the far side of the lobe extension, the intrinsic RM changes from about $+10$ to $-5$~rad~m$^{-2}$.  Toward the southeast, there is a mixture of positive and negative RMs in the lobe. These values are comparable to the uncertainties of the RM model.

\subsubsection{Total magnetic field}
\label{sect:total_B}
\begin{figure}[!htbp]
   \centering
   \includegraphics[width=0.95\textwidth]{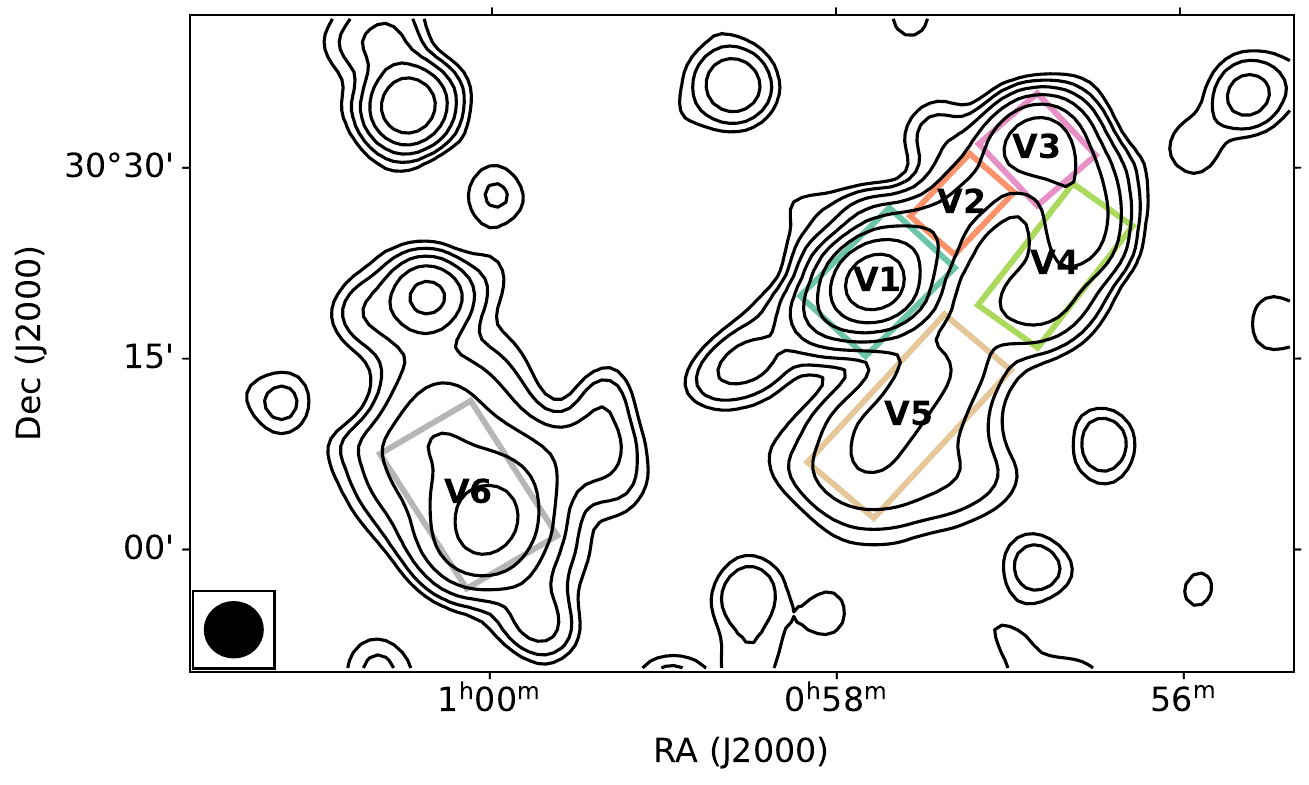}
   \caption{NGC~315 divided into six regions (V1--V6), used to estimate the magnetic field and thermal electron density.}
   \label{fig:B_magentic_mask_regions_map}
\end{figure}

We divided NGC 315 into six regions according to its morphology: core (V1), northwest jet (V2), northwest lobe (V3 and V4), extension of northwest lobe (V5), and southeast lobe (V6) shown in Fig.~\ref{fig:B_magentic_mask_regions_map}. Similar division of these regions was conducted by  \citet{Mack1997}. We then estimated the total magnetic field strength $B$ for each individual region assuming energy equipartition between the magnetic field and cosmic-ray electrons following \citet{Pacholczyk1970}:
\begin{equation}
\label{eq:B}
B = \left( \frac{V \Phi}{6\pi (1+k) c_{12} L} \right)^{-2/7}
\end{equation}
where $V$ is the volume, $\Phi$ is the filling factor for the emitting medium, $k$ is the ratio of the proton energy to the electron energy, $L$ is the luminosity and $c_{12}$ is a constant tabulated by \citet{Pacholczyk1970}. Each region was assumed to be a cylinder with its axis aligned with the jet axis, and its height and diameter of the circular base were derived from the rectangle in Fig.~\ref{fig:B_magentic_mask_regions_map}. The broadening caused by the beam was also corrected. The resulting volumes $V$ are listed in Table~\ref{tab:beq_regions}. The luminosity was estimated from the flux density integrated over the area of each region and over the frequency from 10~MHz to 10~GHz using the spectral index in Fig.~\ref{fig:spc_map}, and the results are also listed in Table~\ref{tab:beq_regions}. We assumed both $\Phi$ and $k$ to be unity and derived the total magnetic field strength $B$ using Eq.~(\ref{eq:B}). We found that the total magnetic field strength is nearly uniform across NGC~315.

The assumptions of a cylinder for $V$ and unity for $\Phi$ and $k$ are standard when estimating magnetic field strength with energy equipartition~\citep{Miley+1980, Mack1998, Beck+2005}. Note that the power index is $-2/7$ in Eq.~(\ref{eq:B}), implying that $B$ depends only weakly on $V$, $\Phi$, and $k$.

%meaning that $B$ is not very sensitive with $V$, $\Phi$, and $k$.

\subsubsection{Thermal electron density}
The thermal electron density $n_e$ of NGC~315 can be estimated from the intrinsic RM (Fig.~\ref{fig:RM_comparison}, bottom panel) as
$$n_e \approx \frac{\rm |RM|}{812B_\parallel\mathcal{L}},$$
where RM is in rad~m$^{-2}$, $B_\parallel$ is the line of sight component of the magnetic field in $\mu$G, $\mathcal{L}$ is the path length along the line of sight in kpc, and $n_e$ is in cm$^{-3}$. For each region, we adopted a median intrinsic RM value. Assuming that the magnetic field is uniform and isotropic, $B_\parallel$ is equal to $B/\sqrt{3}$. We also assumed $\mathcal{L}$ to be equal to the diameter of the cylinders in Sect.~\ref{sect:total_B}. The path length $\mathcal{L}$, RM and estimated $n_e$ are listed in Table~\ref{tab:beq_regions}. Here, we did not take into account the random magnetic field. Therefore, $B_\parallel$ should be regarded as a upper limit and the estimated $n_e$ is thus a lower limit. Given the uncertainties in the source geometry and the foreground RM model,the B and ne values in Table~\ref{tab:beq_regions} are order-of-magnitude estimates.

\begin{table}
\centering
\caption{Parameters for regions V1-V6.}
\label{tab:beq_regions}
\begin{tabular}{lcccccc}
\hline\hline
%\toprule
Region & $V$ (cm$^3$) & $L$ (erg s$^{-1}$) & $B$ ($\mu$G) & $\mathcal{L}$ (kpc) & RM (rad m$^{-2}$) & $n_e$ (cm$^{-3}$) \\
\hline
V1 & 9.3e+70 & 6.7e+40 & 0.9 & 109.2 & 2.5 & 5.2e-5\\
V2 & 2.8e+70 & 5.9e+39 & 0.7 & 24.4 & 6.6 & 8.2e-4\\
V3 & 4.8e+70 & 1.4e+40 & 0.8 & 83.6 & 6.8 & 2.1e-4\\
V4 & 7.7e+70 & 1.0e+40 & 0.6 & 72.9 & 8.0 & 3.8e-4\\
V5 & 1.4e+71 & 7.9e+39 & 0.5 & 104.4 & $-$3.7 & 1.5e-4\\
V6 & 1.7e+71 & 2.7e+40 & 0.6 & 139.3 & $-$0.8 & 1.8e-5\\
\hline
\end{tabular}
\end{table}

The thermal electron density is $\sim10^{-4}\ \mathrm{cm^{-3}}$ in the northwest jet, northwest lobe and its extension and of order $\sim10^{-5}\,\mathrm{cm^{-3}}$ in the core and the southeast lobe. The NGC~315 group contains low-density X-ray-emitting gas \citep{Chen2012}. If the environment of NGC~315 is assumed to have a typical low-density IGM density of about $\sim10^{-5}~\mathrm{cm^{-3}}$ \citep{Klein1996}, the electron density inside NGC~315 is larger than or comparable to that of the environment, suggesting that buoyancy seems not to be the primary explanation for the asymmetric morphology of NGC~315 \citep{Dea2023}. However, this is inconclusive because the estimate of thermal electron density is uncertain due to the large uncertainty of intrinsic RM.

\subsection{A close look at the structure from the core to the extension of the lobe}

\begin{figure}[!htbp]
   \centering
   \includegraphics[width=0.95\textwidth]{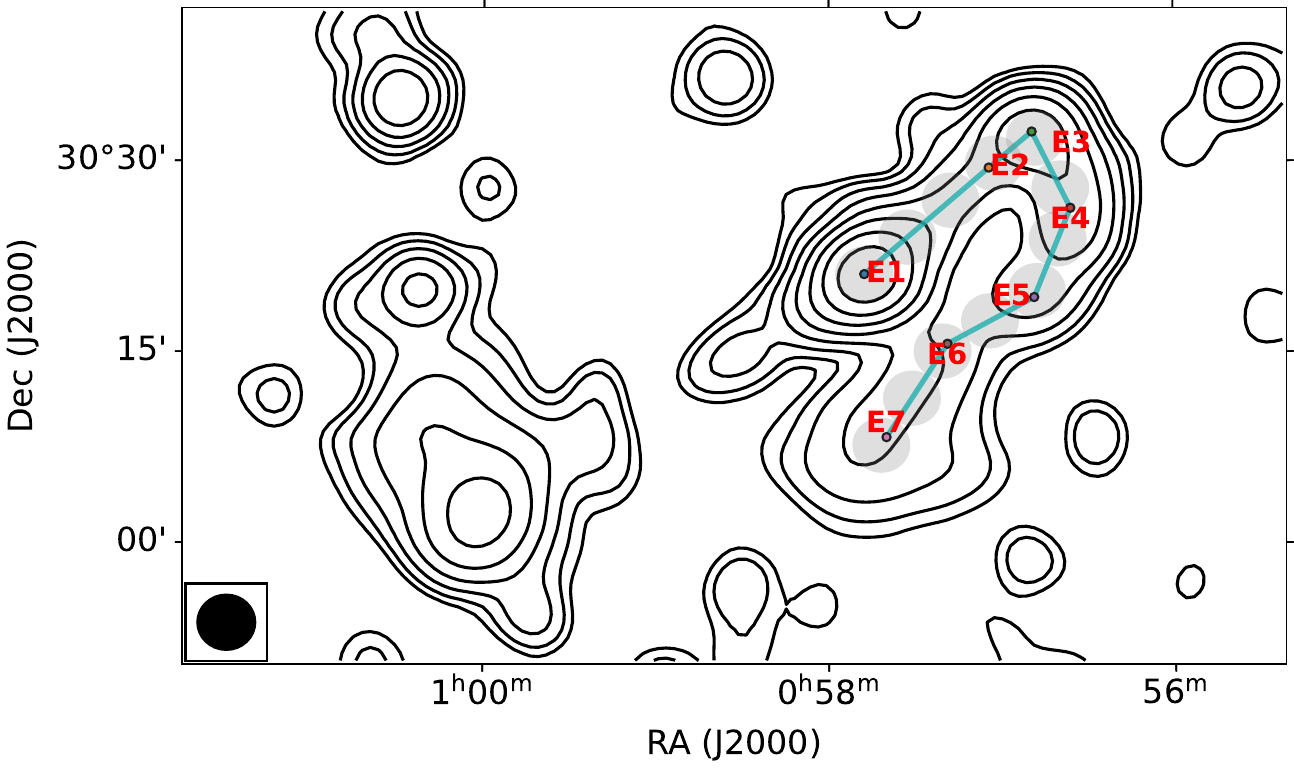}
   \caption{Path from the core to the extension of the northwest lobe marked as a green line. Each shaded circle has a size of $4\farcm35$ and represents a sampling point. Key structures along the path are marked with E1-E7.}
   \label{fig:profile_map}
\end{figure}

We defined a path from the core to the extension of the northwest lobe approximately along the total intensity ridge, shown in Fig.~\ref{fig:profile_map}. The key structures along the path are marked with E1-E7. We then sampled fractional polarization, intrinsic RM, and spectral index along the path in an interval of a common beam size $4\farcm35$ to obtain the profiles. 

\subsubsection{Fractional polarization}
\label{sect:pf_analysis}

\begin{figure}[!htbp]
   \centering
   \includegraphics[width=0.95\textwidth]{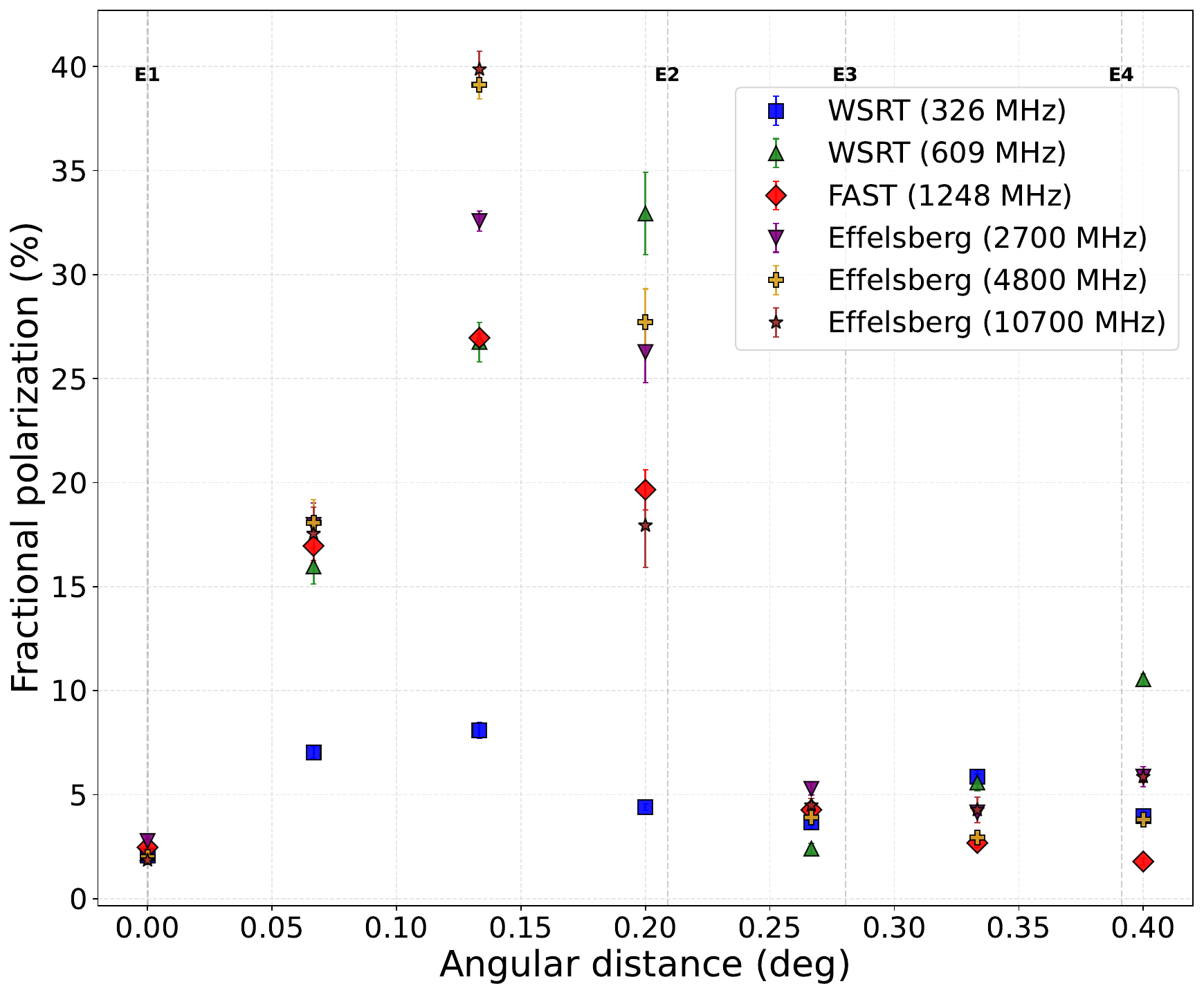}
   \caption{Fractional polarization profiles for different frequencies along the path defined in Fig.~\ref{fig:profile_map}.}
   \label{fig:profile_PF_multi}
\end{figure}

We present fractional polarization profiles in Fig.~\ref{fig:profile_PF_multi}. For fractional polarization, in addition to FAST observations, we also included WSRT observations at 326 and 609 MHz and Effelsberg observations at 2.7, 4.8, and 10.7 GHz~\citep{Mack1997}. All data were convolved to a common resolution of $4\farcm35$ to derive fractional polarization. The error bars were estimated as the standard deviation within each sampling area divided by $\sqrt{N}$, where $N$ is the number of pixels within the area. Because the flux density falls with frequency, the extended emission in E5--E7 is too weak to be detected by Effelsberg at 4.8 and 10.7 GHz. The multi-frequency fractional-polarization profiles are therefore shown only from E1 to E4.

The observed fractional polarization at $\lambda$ can be written as $p(\lambda)=f p_0$, where $f$ is depolarization factor and $p_0$ is the intrinsic fractional polarization. For wavelength-dependent depolarization, $f\propto\sin(2{\rm RM}\lambda^2)/(2{\rm RM}\lambda^2)$ for depth depolarization where synchrotron-emitting and Faraday-rotating medium are uniformly mixed or $f\propto \exp(-2\delta_{\rm RM}^2\lambda^4)$ for beam depolarization where a turbulent Faraday-rotating medium is present within a beam~\citep{Sokoloff+98}. Here, $\delta_{\rm RM}$ is RM scatter. For wavelength-independent depolarization, $f$ is related to the fraction of random magnetic field~\citep{Sokoloff+98}.

The fractional polarization is very low at the level of 2-3\% in the core (E1) and does not depend on the frequency. This means no wavelength-dependent depolarization and the fraction of a regular magnetic field is probably small.

%\begin{figure}[!htbp]
%   \centering
%   \includegraphics[width=0.95\textwidth]{fig/profile_RM.pdf}
%   \caption{Intrinsic RM profile along the path defined in Fig.~\ref{fig:profile_map}.}
%   \label{fig:profile_RM}
%\end{figure}

Along the jet, corresponding to the path between E1 and E2, the fractional polarization is high, reaching 40\% at 4.8 and 10.7~GHz; moving continuously toward the northwest lobe, corresponding to the path between E3 and E4, the fractional polarization has a sharply decrease, down to the level of $\sim5\%$. The decrease of fractional polarization cannot be caused by either depth depolarization or beam depolarization. At 10.7~GHz, an RM of about 1500~rad~m$^{-2}$ would be required for fractional polarization to reduce from 18\% around E2 to 5\% around E3 with depth depolarization, and a $\delta_{\rm RM}$ of 1000~rad~m$^{-2}$ would be required by beam depolarization~\citep{Sokoloff+98}. However, with these values, there would be total depolarization at lower frequencies. Around E3, fractional polarization is similar for all frequencies, indicating that depolarization does not depend on wavelength. It could be that there is enhanced random magnetic field in the emitting region. Note that the magnetic field exhibits loop-like structure, and averaging of the structure across the beam could also cause depolarization. However, fractional polarization around E4 is similar to that around E3, but the magnetic field around E4 does not resemble that around E3. This suggests that the low fractional polarization from E3 to E4 is more likely related to random field.

\begin{figure}[!htbp]
   \centering
   \includegraphics[width=0.95\textwidth]{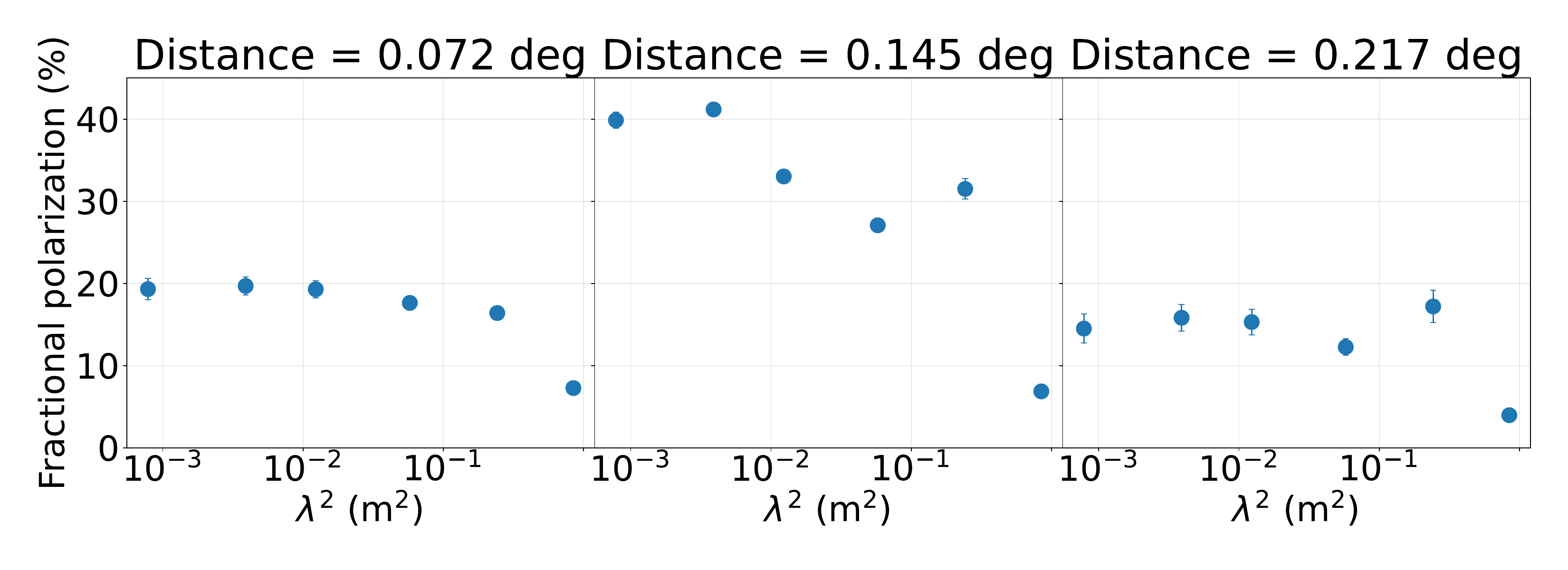}
   \caption{Fractional polarization as a function of $\lambda^2$ for the three sampling points between E1 and E2 along the northwest jet. The frequencies from left to right are 10.7~GHz, 4.8~GHz, 2.7~GHz, 1.248~GHz, 609~MHz and 326~MHz, respectively.}
   \label{fig:PF_lambda2}
\end{figure}

The fractional polarization versus frequency exhibits three kinds of behavior along the northwest jet, as shown in Fig.~\ref{fig:PF_lambda2} where fraction polarization versus wavelength squared ($\lambda^2$) is displayed. 

For the first sampling point (Fig.~\ref{fig:PF_lambda2}, left panel), the polarization fraction is nearly constant at frequencies above 1~GHz, with a clear drop at 326~MHz. For an intrinsic RM of only a few rad~m$^{-2}$ along this path (Table~\ref{tab:beq_regions}), internal Faraday depolarization naturally accounts for the depression at 326~MHz and predicts a noticeable reduction at 609~MHz. The absence of such a drop at 609~MHz may be caused, at least in part, by missing large-scale flux at this frequency (Sect.~\ref{sect:spectral}), which can overestimate $p=P/I$ if the total-intensity structure is more incomplete than the polarized emission.

For the second sampling point (Fig.~\ref{fig:PF_lambda2}, middle panel), the polarization fraction decreases with increasing wavelength. The 609~MHz measurement at this position appears unreliable and is excluded from the following discussion. Assuming depth depolarization, an RM of $\sim45$~rad~m$^{-2}$ is required to account for the decrease from about 40\% at 4.8~GHz to 32\% at 2.7~GHz, whereas only $\sim8$~rad~m$^{-2}$ is required for the decrease from about 32\% at 2.7~GHz to 28\% at 1.248~GHz. A simple uniform slab is therefore inconsistent with the observations. In particular, $\sim45$~rad~m$^{-2}$ substantially exceeds the intrinsic RM along this path (Table~\ref{tab:beq_regions}). It is more likely that the wavelength dependence arises from a stratified Faraday-active medium. In that case, at longer wavelengths, the observed polarized emission may preferentially trace the near side of the source rather than the full line of sight through a single mixed slab. 

For the third sampling point (Fig.~\ref{fig:PF_lambda2}, right panel), the 609~MHz point is likewise unreliable and is excluded. The remaining frequency dependence does not follow a simple Faraday-depolarization sequence. In particular, the fractional polarization at 10.7~GHz is lower than that at the other frequencies, whereas Faraday depolarization alone would suppress longer wavelengths more strongly and leave the highest frequency least affected. An explanation invoking an even larger intrinsic RM, with polarized emission at different frequencies arising from different depths, is not favored. The relatively low $p$ at 10.7~GHz is therefore unlikely to be caused by Faraday depolarization, and its origin remains unclear with the present data.

%\begin{figure}[!htbp]
%    \centering
%    \begin{minipage}{0.32\textwidth}
%        \centering
%        \includegraphics[width=\textwidth]{fig/PD_lambda2_0.067deg.pdf}
%    \end{minipage}
%    \hspace{0.001\textwidth}
%    \begin{minipage}{0.32\textwidth}
%        \centering
%        \includegraphics[width=\textwidth]{fig/PD_lambda2_0.133deg.pdf}
%    \end{minipage}
%    \hspace{0.001\textwidth}
%    \begin{minipage}{0.32\textwidth}
%        \centering
%        \includegraphics[width=\textwidth]{fig/PD_lambda2_0.200deg.pdf}
%    \end{minipage}
%    
%    \caption{Fractional polarization as a function of $\lambda^2$ for the three sampling points between E1 and E2 along the northwest jet.}
%    \label{fig:PF_lambda2}
%\end{figure}

\subsubsection{Spectral index profile}

\begin{figure}[!htbp]
   \centering
   \includegraphics[width=0.95\textwidth]{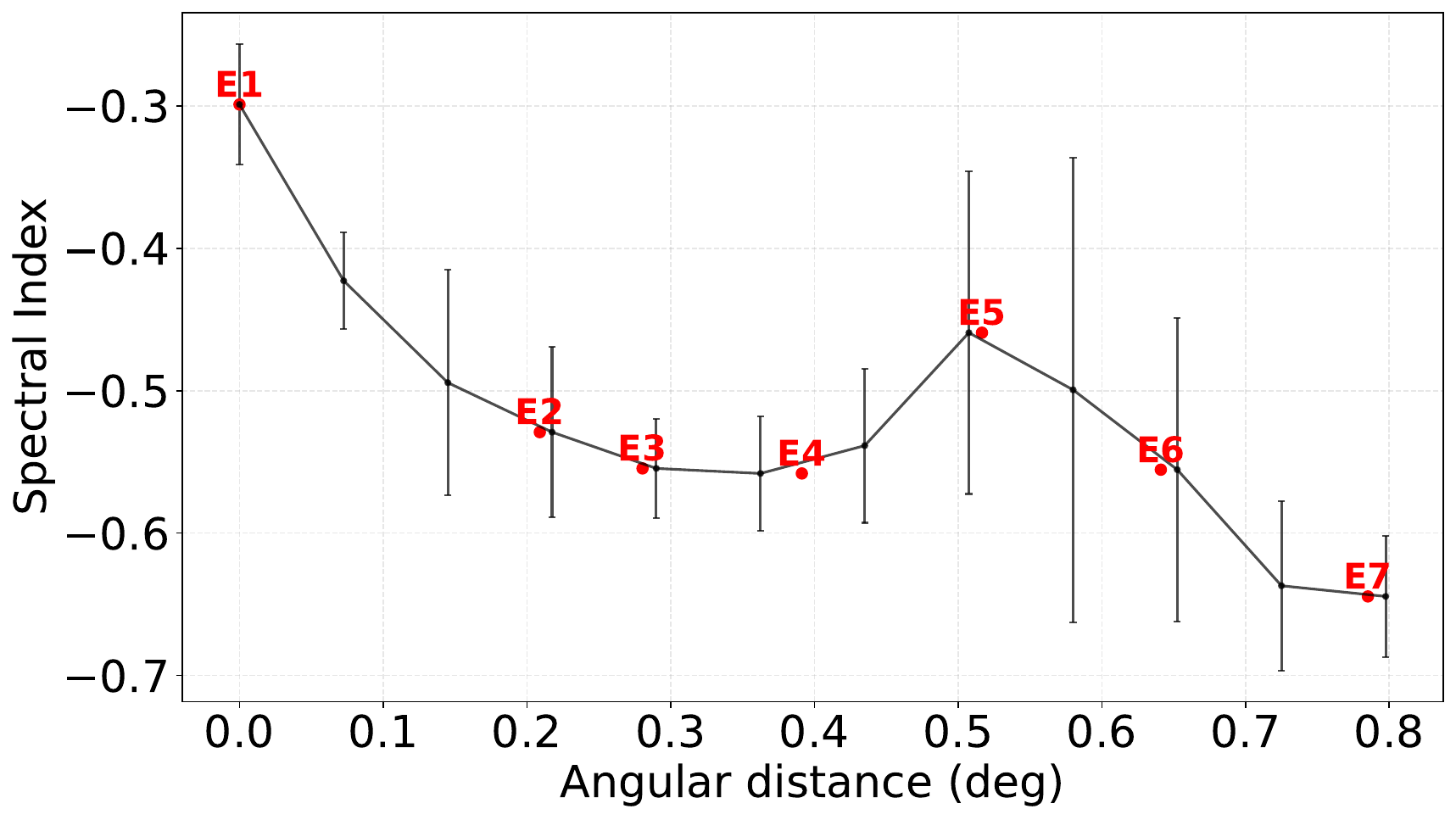}
   \caption{Spectral index profile along the path defined in Fig.~\ref{fig:profile_map}. The sampling points were not necessarily centered on the E1--E7 markers, so the markers may sometimes be offset from the sampling positions. Error bars are median
  $\sqrt{\chi^2/\mathrm{dof}}$--scaled uncertainties from Fig.~\ref{fig:spc_map}}
   \label{fig:profile_spc}
\end{figure}

The spectral index profile is presented in Fig.~\ref{fig:profile_spc}. The error bars were estimated as the median uncertainty within each sampling area from the bottom panel of Fig.~\ref{fig:spc_map}. From the core to the northwest lobe along the jet, corresponding to the path between E1 and E3, the spectral index gradually decreases from $\alpha \approx -0.30$ to $\alpha \approx -0.50$. This is consistent with synchrotron cooling of relativistic electrons propagating in the jet. At the far side (E7) of the extension of the northwest lobe, the spectral index reaches a minimum of $\alpha\approx -0.65$.

\section{Conclusions}
\label{sect:con}
We present new FAST continuum and polarization observations of NGC~315 at 1.0--1.5 GHz and investigate the properties of its large-scale emission. The main results are summarized as follows.

\begin{enumerate}
\item Polarized emission is clearly detected in the southeast lobe, the northwest jet and lobe, and the ``back-flowing'' extension from the northwest lobe, allowing the magnetic-field structure to be traced over a larger spatial extent than in previous studies.

\item We estimated a total magnetic field strength that is less than $\sim1$~$\mu$G across the source based on energy equipartition.

\item We estimate the thermal electron density to be of order $10^{-5}$--$10^{-4}$~cm$^{-3}$, as a model dependent lower limit. A direct comparison with the group
  IGM remains inconclusive because of the large intrinsic RM uncertainty and the
  lack of an NGC 315 specific environmental $n_e$.

\item We suggest the presence of a random magnetic-field component toward the northwest lobe and its extension. We also find that the spectrum steepens from the core toward the outer northwest regions, consistent with radiative cooling.
\end{enumerate}

\normalem
\begin{acknowledgements}
This work is supported by the National Natural Science Foundation of China (No. 12433006) and the National SKA Program of China (Grant No. 2022SKA0120101). W. Z. is supported by the 16th Graduate Research 968 and Innovation Project of Yunnan Province, China (Project No. KC-242410116). This work made use of the data from FAST (Five-hundred-meter Aperture Spherical radio Telescope) (https://cstr.cn/31116.02.FAST). FAST is a Chinese national mega-science facility, operated by National Astronomical Observatories, Chinese Academy of Sciences. We thank the anonymous referee for the comments that helped improve the paper.
\end{acknowledgements}

\appendix
\section{Intensity spectra of selected pixels.}
\begin{figure*}[p]
\centering

\begin{subfigure}{0.32\textwidth}
    \includegraphics[width=\linewidth]{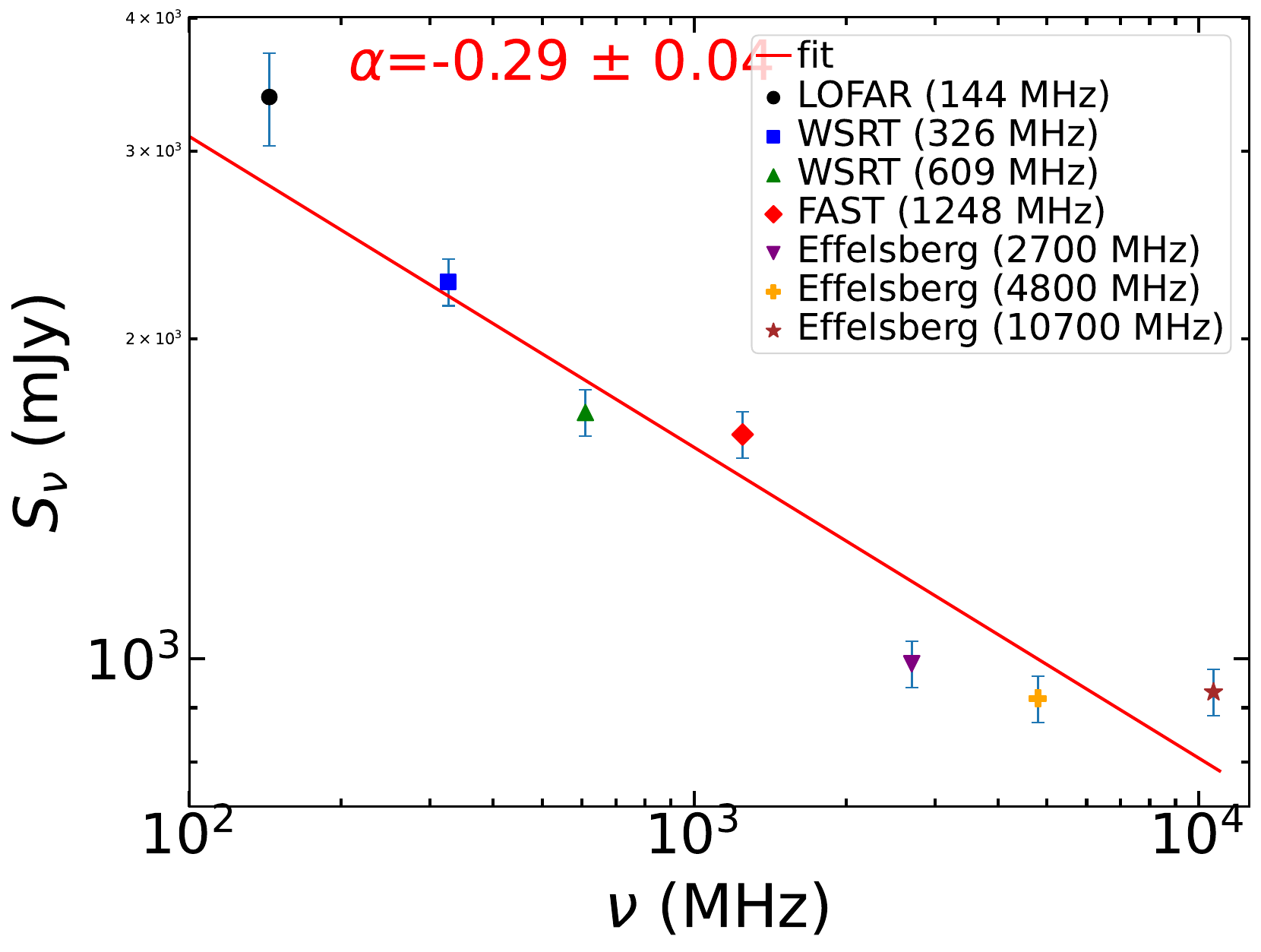}
\end{subfigure}
\begin{subfigure}{0.32\textwidth}
    \includegraphics[width=\linewidth]{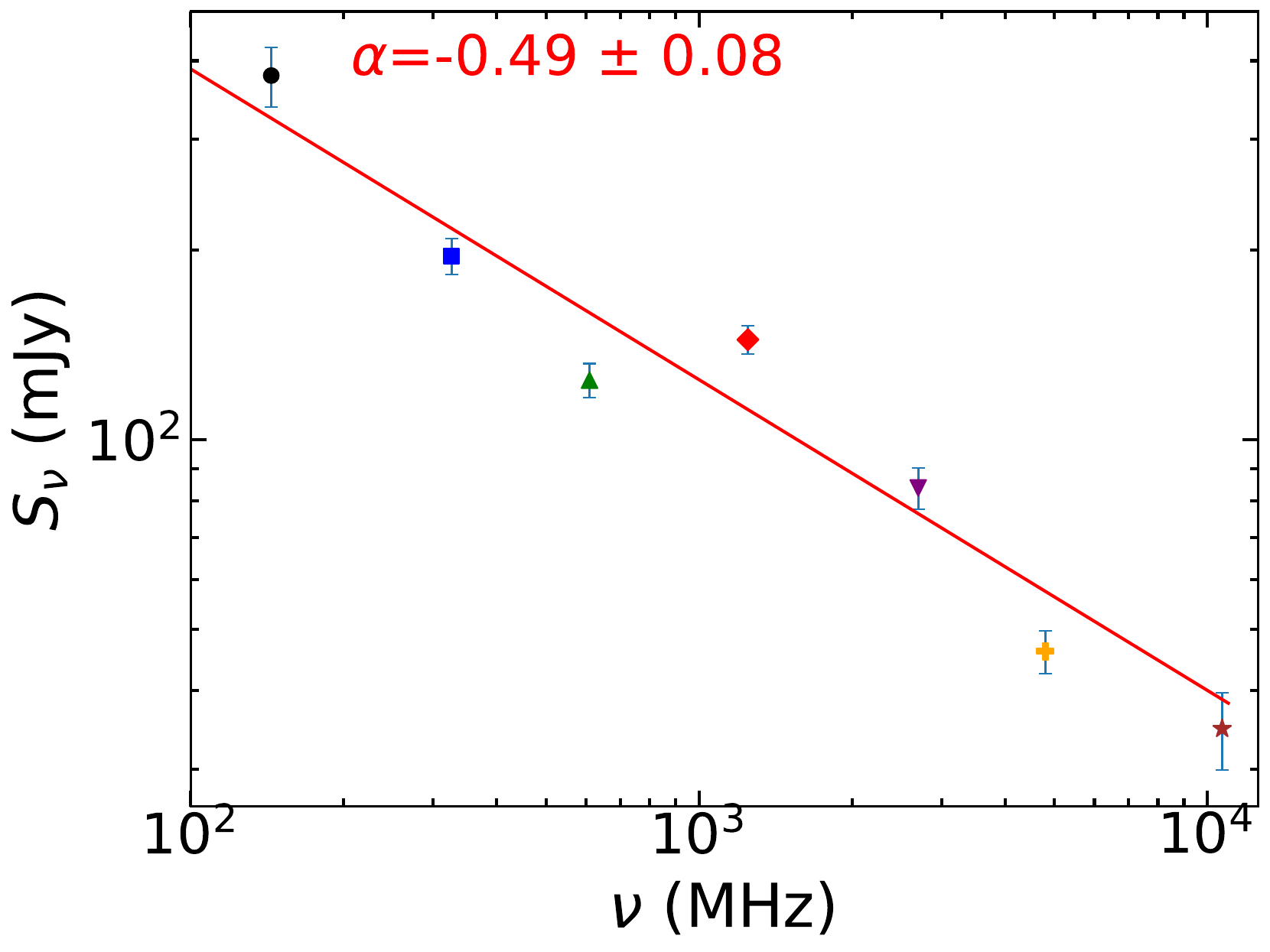}
\end{subfigure}
\begin{subfigure}{0.32\textwidth}
    \includegraphics[width=\linewidth]{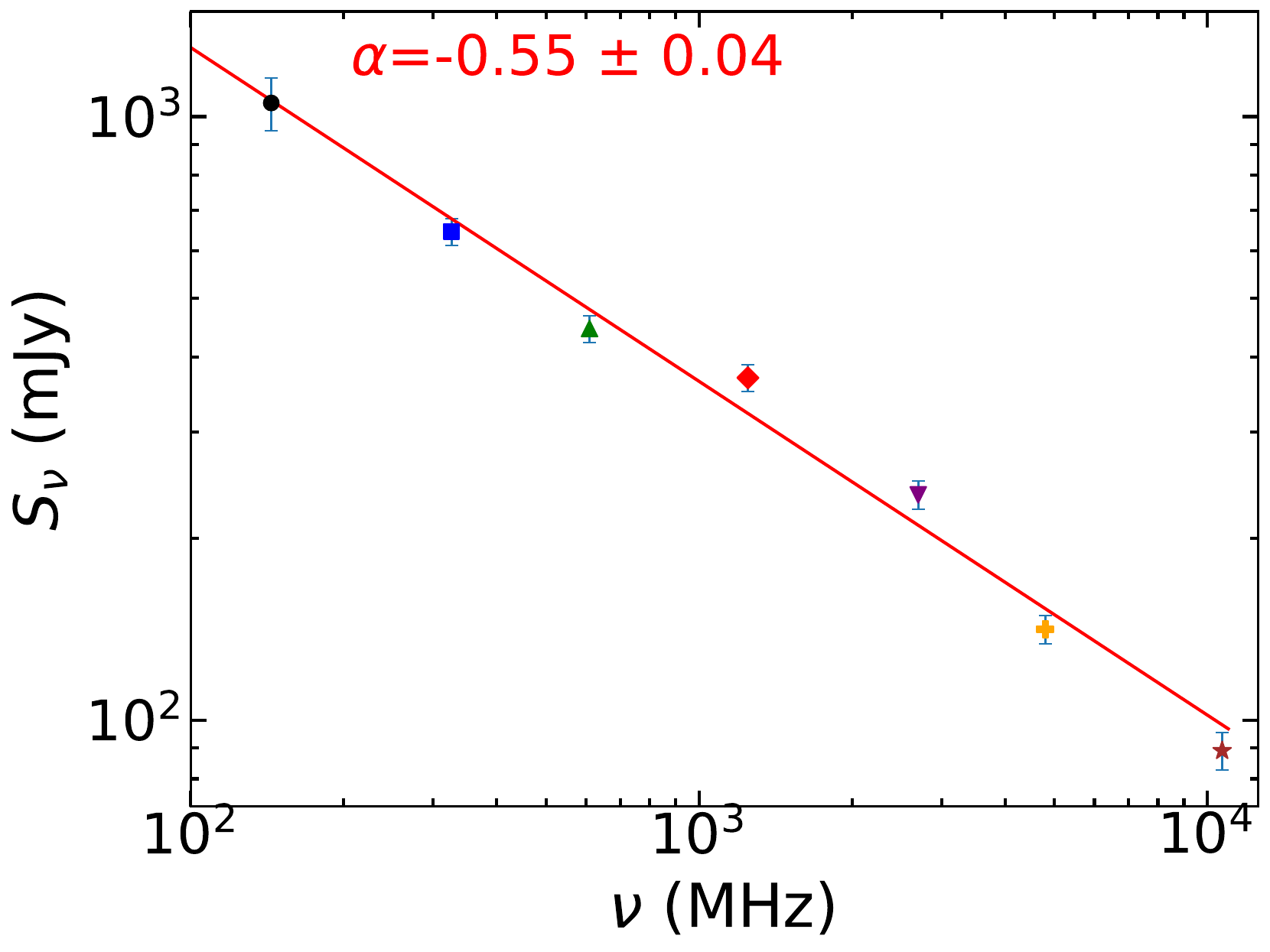}
\end{subfigure}

\vspace{0.4em}

\begin{subfigure}{0.32\textwidth}
    \includegraphics[width=\linewidth]{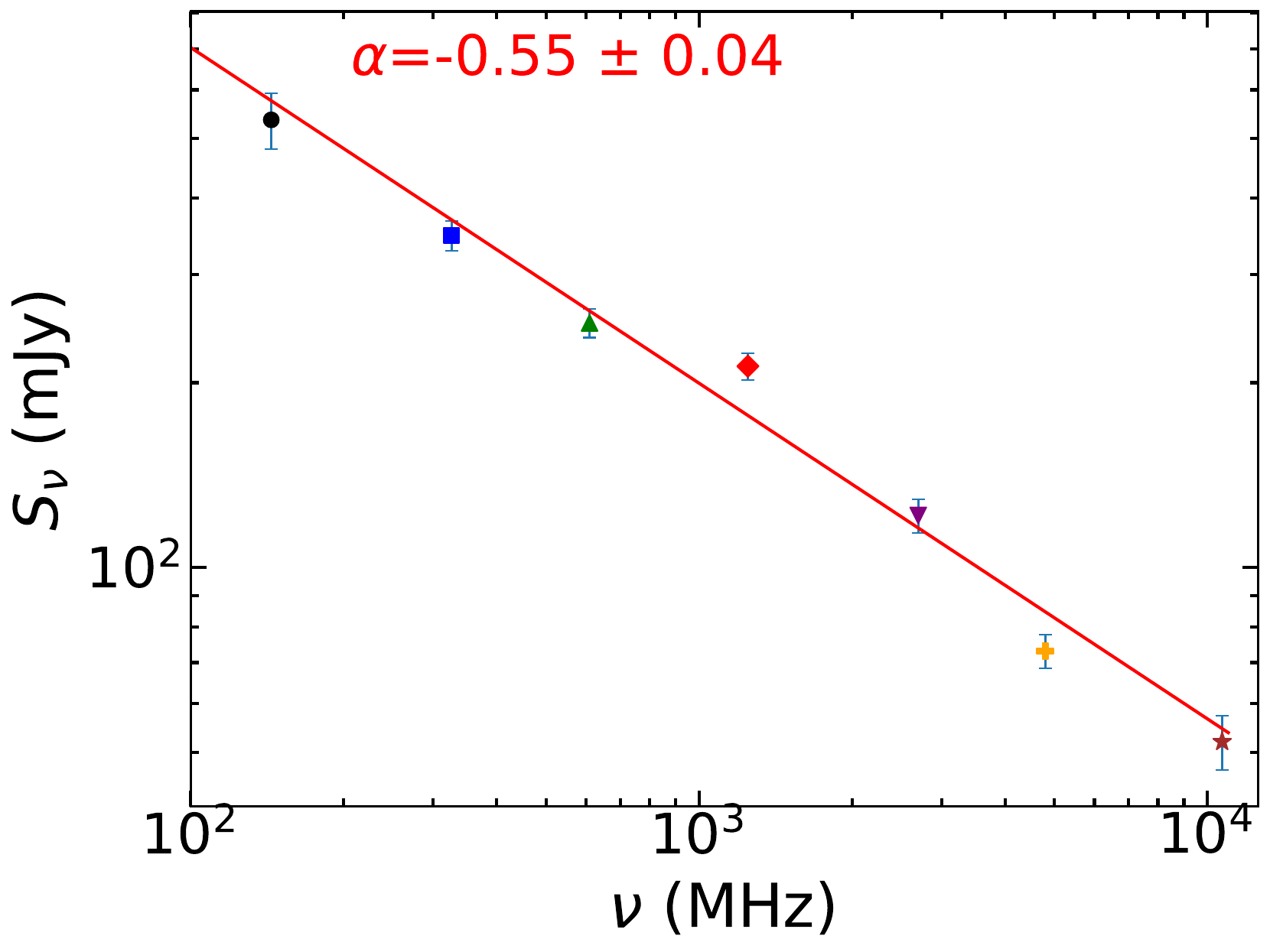}
\end{subfigure}
\begin{subfigure}{0.32\textwidth}
    \includegraphics[width=\linewidth]{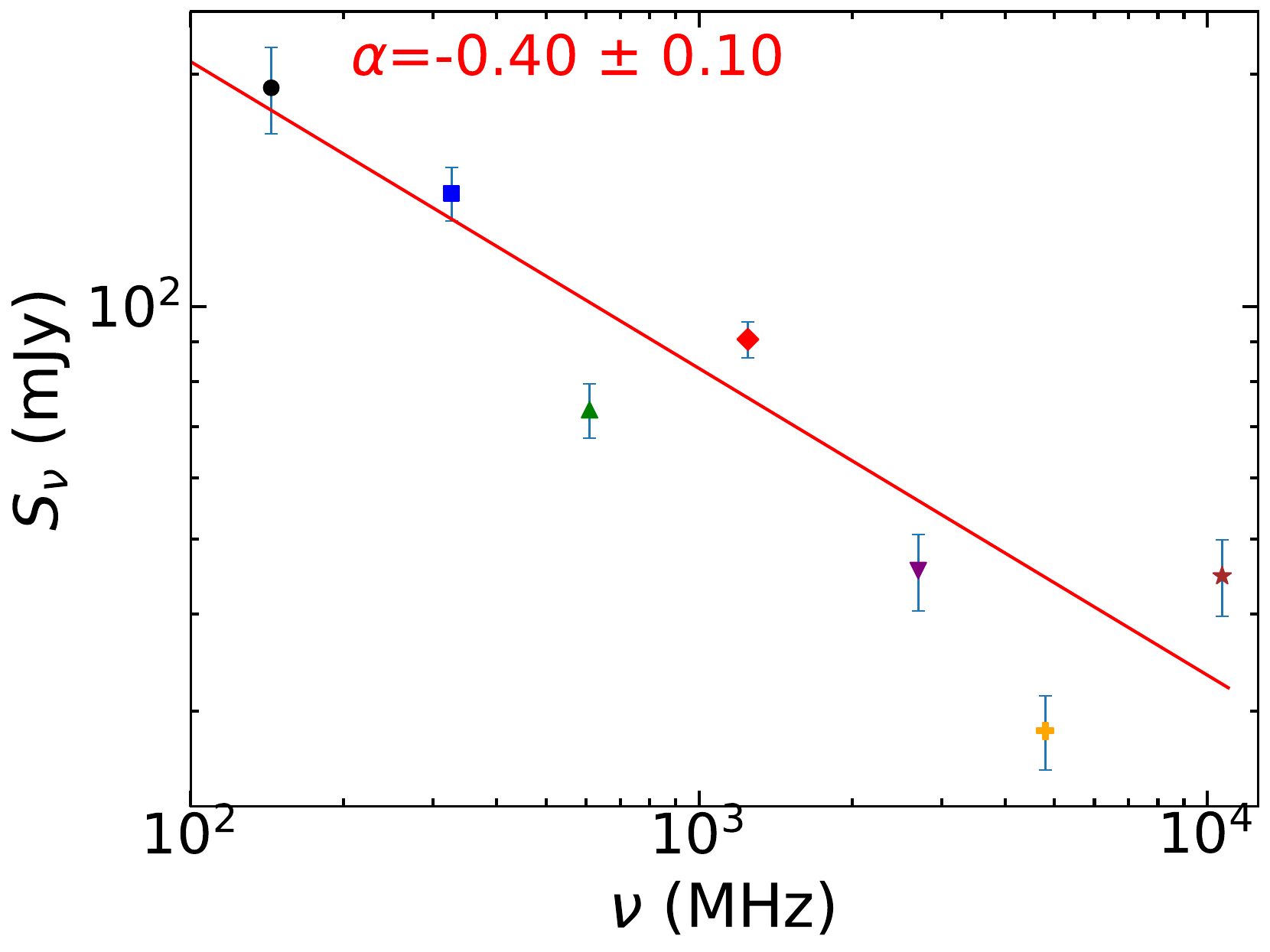}
\end{subfigure}
\begin{subfigure}{0.32\textwidth}
    \includegraphics[width=\linewidth]{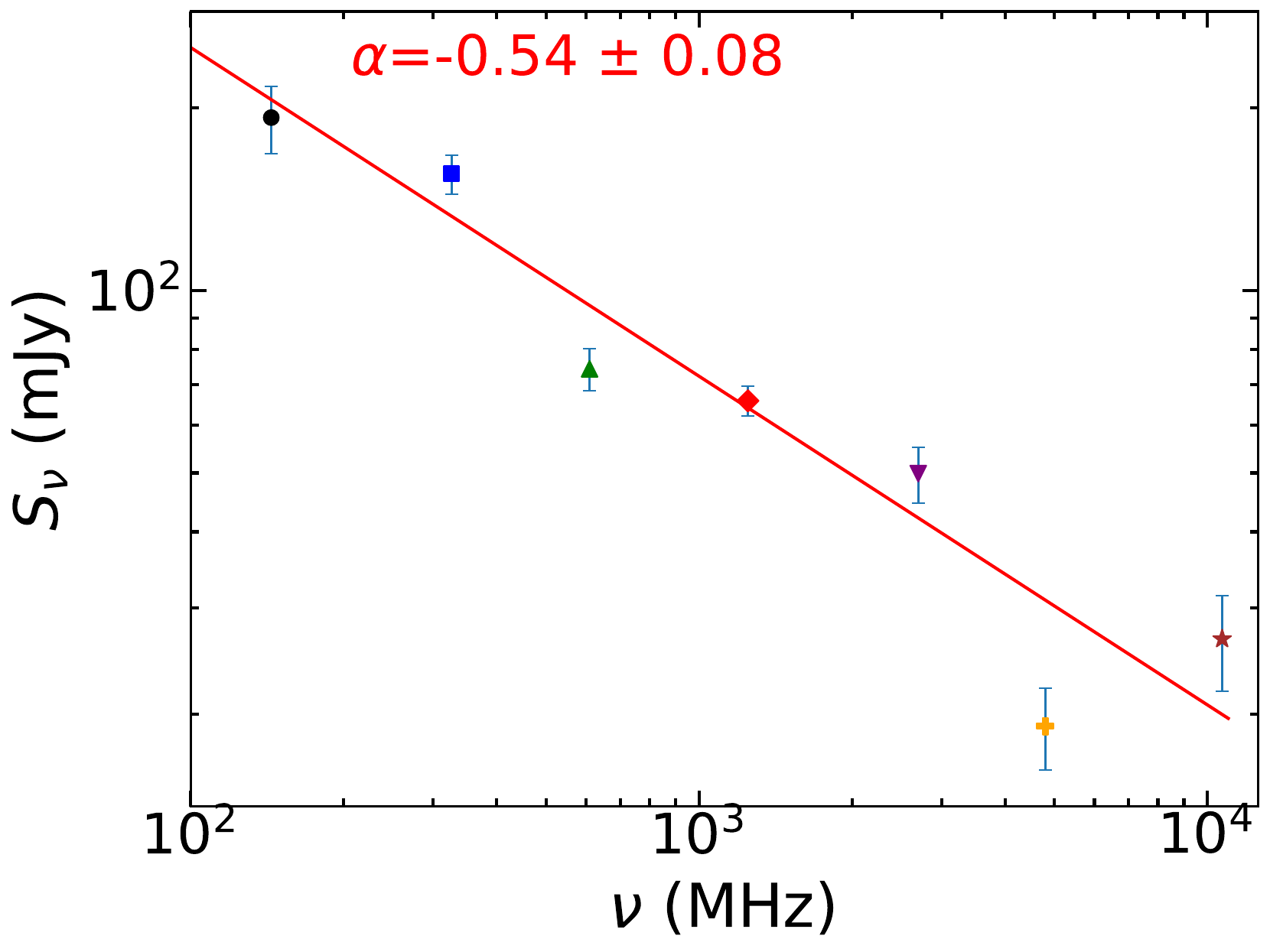}
\end{subfigure}

\vspace{0.4em}

\begin{subfigure}{0.32\textwidth}
    \includegraphics[width=\linewidth]{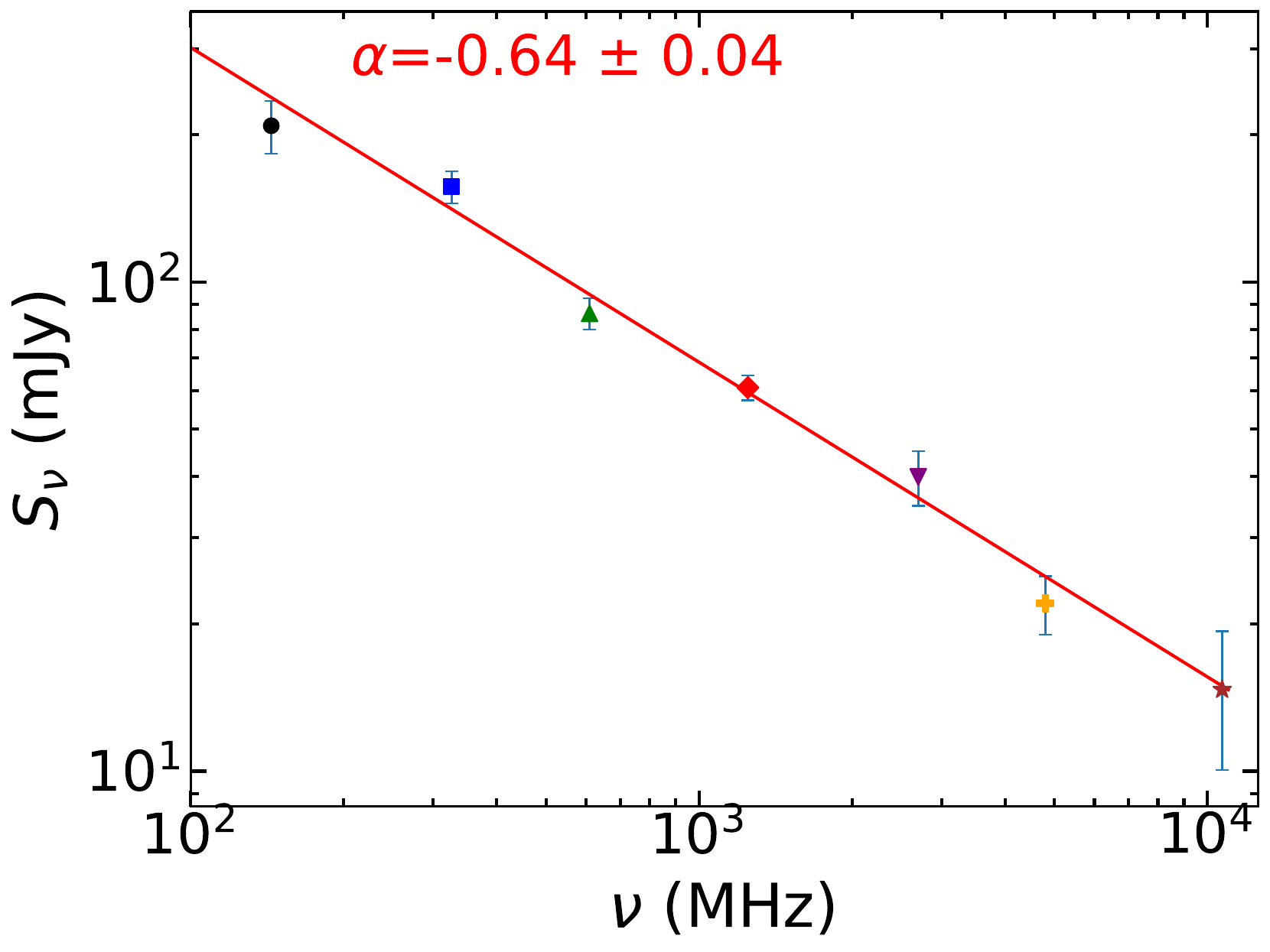}
\end{subfigure}
\begin{subfigure}{0.32\textwidth}
    \includegraphics[width=\linewidth]{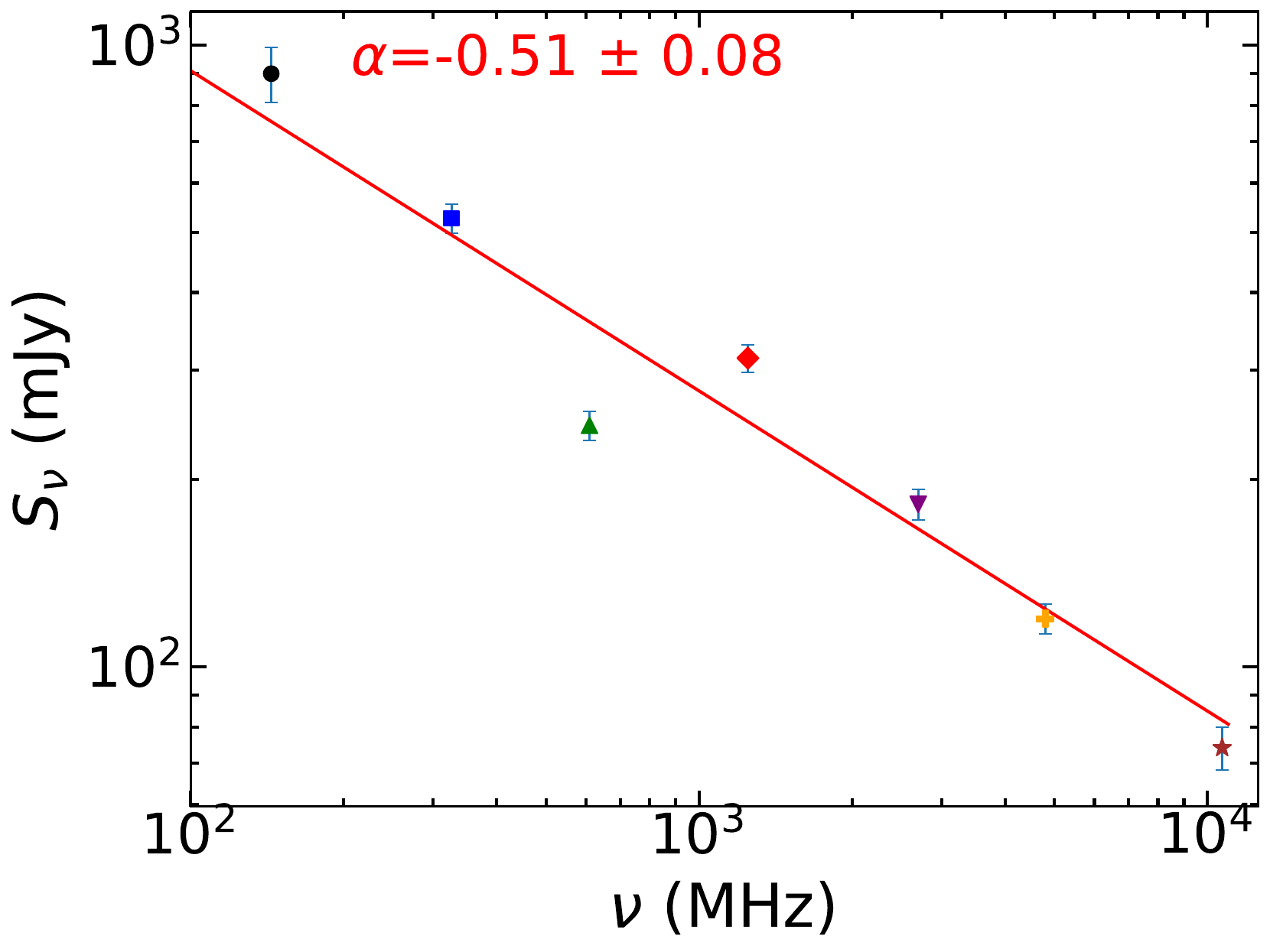}
\end{subfigure}
\begin{subfigure}{0.32\textwidth}
    \includegraphics[width=\linewidth]{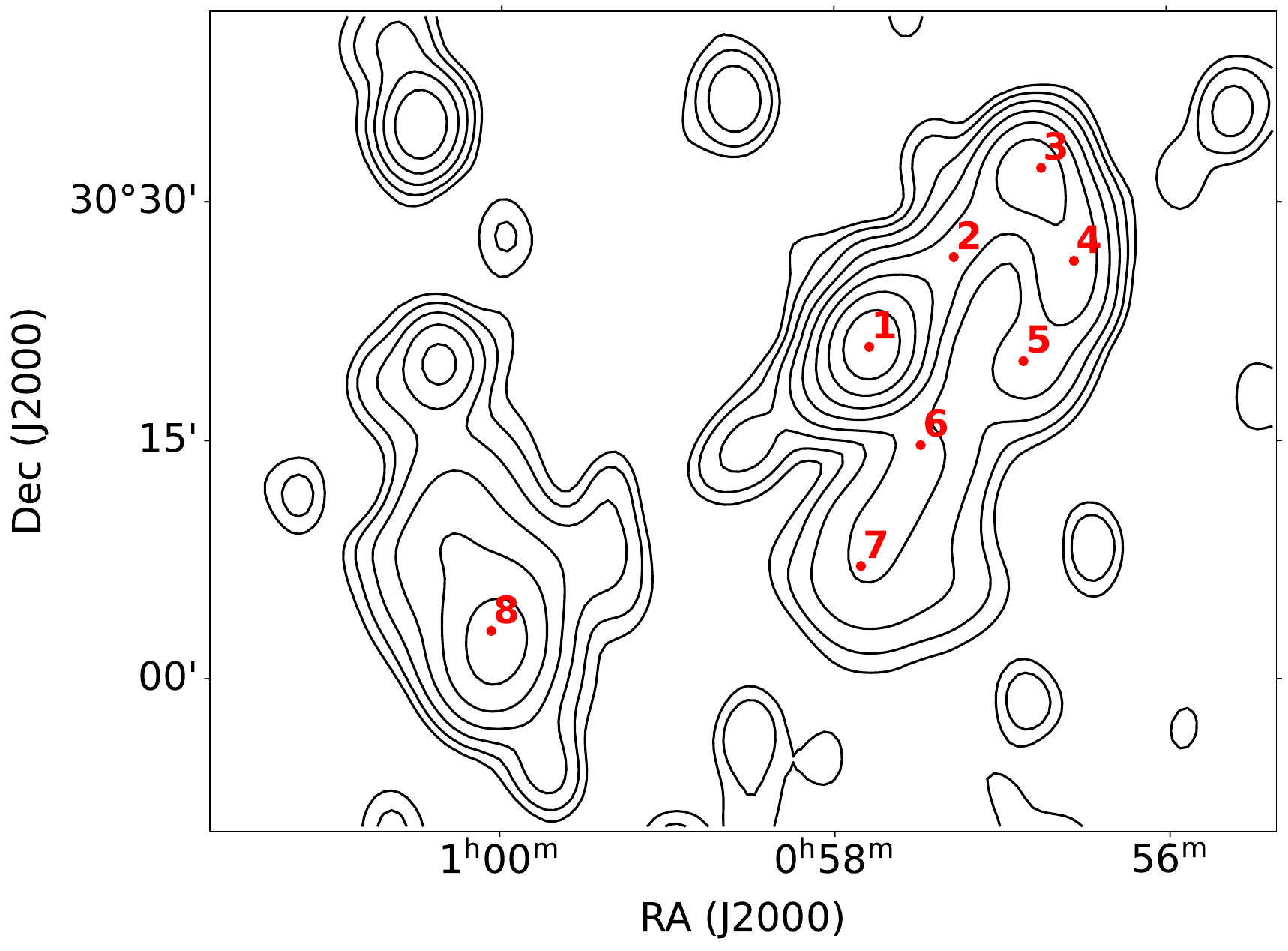}
\end{subfigure}

\caption{Total intensity spectra for selected pixels. From left to right and from top to bottom, the panels correspond to the regions labeled 1-8 in the contour map at the bottom right panel. In each panel, the red line shows the fitted power-law spectrum and measurements at different frequencies are displayed using different symbols and colors.}
\label{fig:appendix_spectra}
\end{figure*}

\clearpage
\bibliographystyle{raa}
\bibliography{bibtex}

\end{document}